\documentclass[nofootinbib,preprintnumbers,floatfix,onecolumn,superscriptaddress,a4paper]{revtex4}

\usepackage{amsmath}
\usepackage{amssymb}
\usepackage{graphicx}
\usepackage{xcolor}
\usepackage{soul}

\usepackage[T1]{fontenc} 
\usepackage{slashed}
\usepackage[colorlinks=true,citecolor=blue,linkcolor=purple,urlcolor=purple]{hyperref}
\usepackage{color} 

\newcommand{\id}{\mathbb{I}}

\def\gsim{\raise0.3ex\hbox{$\;>$\kern-0.75em\raise-1.1ex\hbox{$\sim\;$}}}
\def\lsim{\raise0.3ex\hbox{$\;<$\kern-0.75em\raise-1.1ex\hbox{$\sim\;$}}}

\newcommand{\AddrIFIC}{%
Instituto de F\'{\i}sica Corpuscular (CSIC-Universitat de Val\`{e}ncia),
Apdo. 22085, E-46071 Valencia, Spain
}

\begin{document}

\preprint{IFIC/20-08}

\title{Long-lived charged particles and multi-lepton signatures 
from neutrino mass models}

\author{Carolina Arbel\'aez R}
\email{carolina.arbelaez@usm.cl}
\affiliation{Universidad T\'ecnica Federico Santa Mar\'ia and
Centro Cient\'ifico Tecnol\'ogico de Valparaiso CCTVal
Casilla 110-V, Valparaiso, Chile}

\author{Giovanna Cottin}
\email{giovanna.cottin@uai.cl}
\affiliation{Departamento de Ciencias, Facultad de Artes Liberales, 
Universidad Adolfo Ib\'a\~{n}ez,
Diagonal Las Torres 2640, Santiago, Chile}

\author{Juan Carlos Helo}
\email{jchelo@userena.cl}
\affiliation{Departamento de F\'{i}sica,
Facultad de Ciencias, Universidad de La Serena,\\
Avenida Cisternas 1200, La Serena, Chile}

\author{Martin Hirsch} \email{mahirsch@ific.uv.es }
\affiliation{\AddrIFIC}

\begin{abstract}

Lepton number violation (LNV) is usually searched for by the LHC
collaborations using the same-sign di-lepton plus jet signature.  In
this paper we discuss multi-lepton signals of LNV that can arise with
experimentally interesting rates in certain loop models of neutrino
mass generation. Interestingly, in such models the observed smallness of the
active neutrino masses, together with the high-multiplicity of the
final states, leads in large parts of the viable parameter space of
such models to the prediction of long-lived {\em charged} particles,
that leave highly ionizing tracks in the detectors. We focus on one
particular 1-loop neutrino mass model in this class and discuss its
LHC phenomenology in some detail.

\end{abstract}

\maketitle

\section{Introduction}

Lepton number violation (LNV) is usually searched for by the  LHC
collaborations using the same-sign di-lepton plus jet signature
\cite{Aaboud:2018spl,Sirunyan:2018pom}.  This signal was first
proposed in the context of the left-right symmetric model
\cite{Keung:1983uu}, but appears -- at least in principle -- in all
Majorana neutrino mass models.\footnote{Signal rates and kinematics of
  this final state are of course highly model dependent.}  In this
paper, we discuss multi-lepton signals, possibly accompanied by
long-lived heavy particles leaving charged tracks in the LHC
detectors. Here, by ``multi-lepton'' we understand final states with
4-, 6- or even more leptons accompanied by jets, but {\em without
  missing energy}.  As we will show, this kind of exotic LNV
signatures appears in specific loop models of neutrino mass generation.

From the theoretical point of view, the smallness of neutrino masses
could be understood, if neutrino masses are generated radiatively.
Radiative neutrino mass models have a long history
\cite{Zee:1980ai,Cheng:1980qt,Zee:1985id,Babu:1988ki}, see also the
recent review \cite{Cai:2017jrq}. Different radiative models can be
classified according to the topology of the diagram from which the
neutrino mass is generated. Systematic classifications have been done
for 1-loop \cite{Bonnet:2012kz}, 2-loop \cite{Sierra:2014rxa} and even
3-loop \cite{Cepedello:2018rfh} models.

\begin{figure}
\begin{centering}
\includegraphics[scale=0.6]{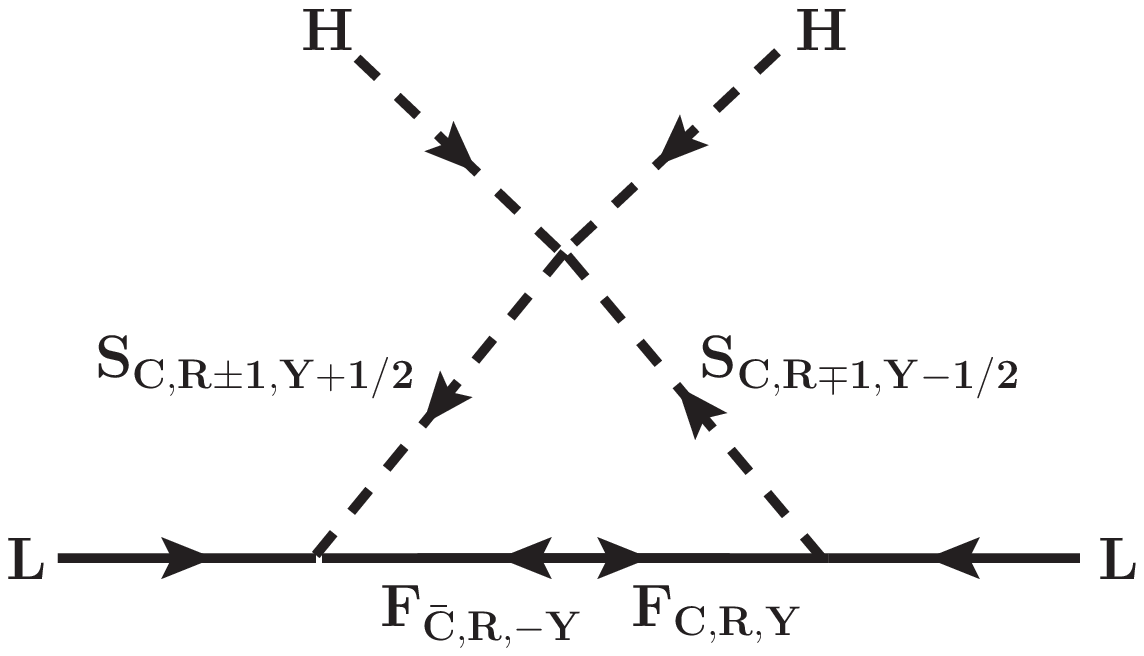}\hskip15mm
\includegraphics[scale=0.6]{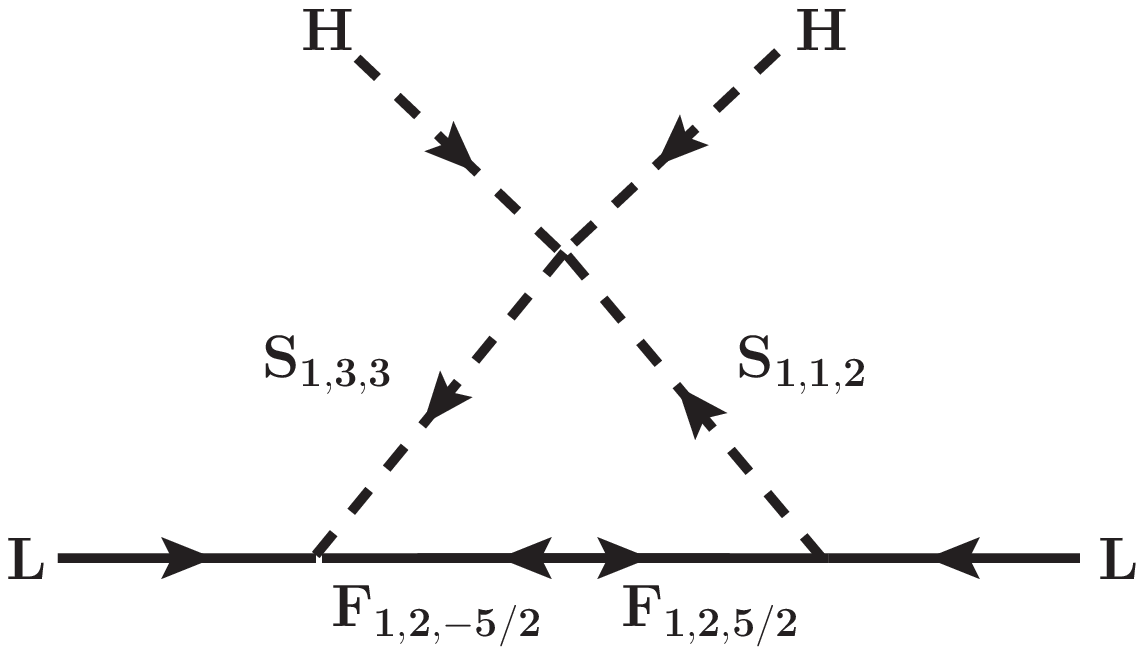}
\end{centering}
\protect\caption{\label{fig:Mdl1} To the left: The general 1-loop $d=5$
  neutrino diagram $T_3$ \cite{Bonnet:2012kz}. To the right: Choosing
  $C=1$, $R=2$ and $Y=5/2$ results in the prototype model that
  predicts the multi-lepton LNV signals discussed in the text.}
\end{figure}

Customarily, when constructing neutrino mass models, model builders
use the smallest available representations to do required tasks.  For
example, in fig. (\ref{fig:Mdl1}) to the left we show one particular
1-loop $d=5$ diagram, the only diagram from topology $T_3$ in
\cite{Bonnet:2012kz}.  
Choosing $C=1$, $R=1$ and $Y=0$ results in the {\em scotogenic} model
\cite{Ma:2006km}. This renowned neutrino mass model requires an
additional $Z_2$ symmetry, under which the particles internal to the
loop are odd, in order to avoid the tree-level type-I seesaw. But once
we accept this premise, as a free bonus the scotogenic model contains
two potential dark matter candidates. It should be clear, however,
that a priori there is no fundamental principle that fixes ($C,R,Y$) 
in such kind of model building exercise.

Choosing instead $C=1$, $R=2$ and $Y=5/2$ results in the particular
model shown to the right. Different from the scotogenic model, this
model variant does not need any additional symmetry (on top of the
standard model gauge group) to avoid tree-level neutrino masses,
i.e. the diagram shown is automatically the leading contribution to
the neutrino mass matrix. More interesting for us, however, is that
this model represents the proto-type for a class of neutrino mass
models with multi-lepton signals at colliders, which are  the subject 
of this paper.

Searches for LNV require signatures with {\em charged} leptons in the
final state. In models based on the type-I
seesaw~\cite{Minkowski:1977sc,Mohapatra:1979ia,Schechter:1980gr},
mixing between the singlet neutrinos,\footnote{$N$ appear under
  various names in the literature: sterile neutrinos, heavy neutrinos
  or also heavy neutral leptons (HNLs).} $N$, and the active neutrinos
of the standard mdoel (SM) leads to a non-zero production cross
section of these states at the LHC: $pp \rightarrow (W^{\pm})^*
\rightarrow l^{\pm} N $. The $N$, being a Majorana particle, can decay
via $N \rightarrow l^{\pm} q\bar{q}$, leading to the standard $2l+ jj$
signal, with both same-sign and opposite-sign leptons.  CMS has
performed a search for Majorana $N$'s using this channel
\cite{Sirunyan:2018xiv}.  The $N$ can also decay as $N \rightarrow
l^{\pm}l'^{\mp}\nu$, which leads to a tri-lepton signature with
missing momentum in the event. CMS has used this signature to search
for seesaw type-I HNLs \cite{Sirunyan:2018mtv}, as well as for the
type-III seesaw \cite{Sirunyan:2019bgz}. While interesting limits on
new particles have been established by these searches, we need to
stress, however, that events with missing energy can {\em never
  establish the existence of LNV}.

We define LNV {\em multi-lepton} events as final states with at least
four charged leptons in the final state (with a lepton number balance
of $L=\pm 2$). Consider the Feynman diagram of a process leading to
$l^{\pm}l^{\pm}jj$. It is easy to add either a photon or a $Z^0$ boson
to this diagram. Decaying $\gamma/Z^0 \to l^+l^-$ will lead to events
such as $l^{\pm}l^{\pm}(l^+l^-)jj$, i.e. a LNV multi-lepton event in
our definition.  However, usually the rates for such a final state are
suppressed relative to $l^{\pm}l^{\pm}jj$ by at least a factor of
$\alpha$ and thus, experimentally uninteresting. This is different in
the models discussed in this paper. In the model defined by
the particles shown in fig. (\ref{fig:Mdl1}) right, multi-lepton
events have actually {\em larger rates} than di-lepton events, as we
will show below. Since such high multiplicity final states should also
have less background at the LHC, we believe a dedicated search for
this kind of signals at the LHC might be worthwhile.

The purpose of this paper is to study in some detail the phenomenology
of the model shown in fig. (\ref{fig:Mdl1}). Since the model uses only
non-singlet states, all beyond SM particles can in principle be
produced at the LHC.  Final states contain always at least two
same-sign leptons and the model predicts final states with up to eight
leptons. The model can easily fit neutrino data. Given the smallness 
of neutrino masses, one expects that also the decay widths of the 
exotic scalars are small. As we will show, this is especially so 
in the parts of the parameter space where $S_{1,3,3}$ is the lightest 
exotic particle. If this is the case, we expect the scalars
from $S_{1,3,3}$ to be long-lived and leave charged tracks in the 
detectors before decaying.

The rest of this paper is structured as follows. In section
\ref{sect:model} we will present the model and give some approximation
formulas for the decay rates of the exotic particles. In section
\ref{sect:res} we discuss our numerical results. In section
\ref{sect:len} we will show decay lengths of the exotic particles of
our model, concentrating mostly on the scalars. In section
\ref{subsect:Cross} we calculate cross sections and discuss the reach
of the LHC for both, present and future luminsities.  We then close
with a more general discussion of multi-lepton signatures.

\section{The Model}
\label{sect:model}

For our numerical results, we will concentrate on a specific model
variant, the neutrino mass diagram is shown in fig. (\ref{fig:Mdl1}).
The model introduces a vector-like fermion pair, $(F,{\bar F})\equiv
(F_{1,2,5/2},F_{1,2,-5/2})$. Here and everywhere else in this paper,
subscripts in the gauge basis denote the transformation properties and
charge under the SM gauge group in the order $SU(3)_C\times
SU(2)_L\times U(1)_Y$.  In the numerical implementation we use three
copies of $(F,{\bar F})$, one for each lepton generation.  We note,
however, that neutrino data could already be explained with just one
copy of $(F,{\bar F})$. The model also has two new scalars, $S_1
\equiv S_{1,1,2}$ and $S_3 \equiv S_{1,3,3}$.

The lagrangian of the model is given by:
\begin{eqnarray}\label{eq:lag}
  {\cal L} = {\cal L}_{SM} &-& m_F F{\bar F}
  - h_{ee} e^c e^c S_{1}^{\dagger} - h_F L FS_{1}^{\dagger}
  - h_{\bar F} L {\bar F}S_{3}
  \\ \nonumber
  & - & m_{S_1}^2 |S_{1}|^2 - m_{S_3}^2 |S_{3}|^2+ \lambda_2 |H|^2 |S_{1}|^2 
  \\ \nonumber
  &+&  \Big(\lambda_{3a}|H^{\dagger}H| |S^{\dagger}_{3}S_{3}|
  + \lambda_{3b} |H S_{3} H^{\dagger}S_{3}^{\dagger}|\Big) 
  + \lambda_4 |S_1|^4 - \lambda_5 HHS_2 S_3^{\dagger}  \\ \nonumber
  & +&  \Big( \lambda_{6a} |S_3^{\dagger}S_3|^2  +\lambda_{6b}
  |S_3S_3 S_3^{\dagger}S_3^{\dagger}|\Big)
  +\lambda_7 |S_1|^2|S_3|^2
\end{eqnarray}


The terms proportional to $\lambda_3$ and $\lambda_6$ have two
independent ${ SU(2)_L}$ contractions each. We used {\texttt Sym2Int}
\cite{Fonseca:2011sy,Fonseca:2017lem} to find all terms in
eq. (\ref{eq:lag}). For the terms in the first line, as well as 
for $\lambda_5$, one has to add the hermitian conjugates as well. 
All Yukawas are to be understood as ($3,3$) matrices in 
generation space.

Note, that for $\lambda_5 \to 0$ lepton number is restored in the
model. In this limit also neutrino masses vanish, so it is not
alllowed phenomenologically. However, since $\lambda_5 \to 0$
corresponds to a symmetry (lepton number conservation), small values
of $\lambda_5$ are technically natural in the sense of t'Hooft.  We
also point out the interaction proportional to $h_{ee}$: This Yukawa
coupling does not appear in the diagram
fig. (\ref{fig:Mdl1}). However, all decays of our exotic particles
must contain this coupling, we therefore call it loosely the ``exit'',
since for $\forall (h_{ee})_{ij} \equiv 0$ the lightest of the loop
particles would be stable. Again, this limit is not allowed
phenomenologically.

After electro-weak symmetry breaking, the doubly charged components 
in $S_1$ and $S_3$ mix. At tree-level their mass matrix is given
by:
\begin{equation}
  M^{2}_{2+}=\begin{pmatrix}
  m_{S_1}^2 - \frac{1}{2}\lambda_2 v^2 &  \frac{1}{2}\lambda_5 v^2 \\
   \frac{1}{2}\lambda_5 v^2 & m_{S_3}^2 - \frac{1}{2}\lambda_{3a} v^2
    \end{pmatrix} .
    \label{eq:mspp}
\end{equation}
This matrix can easily be diagonalized analytically. The mixing
angle can be expressed as:
\begin{equation}
  \tan 2\theta =
  \frac{\lambda_5 v^2}{(m_{S_3}^2 - m_{S_1}^2) 
- \frac{1}{2}(\lambda_{3a} -\lambda_2) v^2} .
  \label{eq:t2t}
\end{equation}
We will call the mass eigenstates $H^{2+}_i$, with $i=1,2$, the symbol
$S$ is reserved for gauge states. Also the mass eigenstates $H^{3+}$
and $H^{4+}$ receive a contribution to their mass proportional to the
Higgs vev:
\begin{eqnarray}\label{eq:msmp}
  m_{4+}^2 = m_{S_3}^2 - \frac{1}{2} ( \lambda_{3a}+\lambda_{3b})v^2,
  \\ \nonumber
   m_{3+}^2 = m_{S_3}^2 - (\frac{1}{2}\lambda_{3a}+\frac{1}{4}\lambda_{3b})v^2.
\end{eqnarray}
As eq. (\ref{eq:msmp}) shows, $m_{4+}^2 < m_{3+}^2$ if $\lambda_{3b} >
0$. Note that the signs of the $\lambda_i$ can be chosen freely. 
For values of $m_{S_3}$ much larger than $v$ and/or for
$\lambda_i\ll 1$ one expects that the states $H^{3+}$, $H^{4+}$ and
one of the $H^{2+}$ will be nearly degenerate.

The neutrino mass matrix in this model can be written as:
\begin{equation}\label{eq:NuMat}
m_{\nu} = {\cal F} \times ( h_{\bar F}^T {\cal M} h_F + h_F^T {\cal M}  h_{\bar F})
\end{equation}
Here, ${\cal F} \simeq 1/(32 \pi^2)\sin(2\theta)$ and ${\cal M}$ is 
a ($3,3$) matrix with diagonal entries given by:
\begin{equation}\label{eq:Mat}
{\cal M}_{ii} = m_{F_i}\Delta B_0(x_1^2,x_2^2),
\end{equation}
where $x_{1,2}=m_{2+,(1,2)}/m_{F_i}$ and 
\begin{equation}\label{eq:delB}
\Delta B(x, y) = x\frac{\log(x)}{(x - 1)} - y\frac{\log(y)}{(y - 1)}.
\end{equation}
As eq.(\ref{eq:NuMat}) shows, the neutrino masses depend on the 
product of two Yukawa couplings, which in principle can be different. 
A complete fit to neutrino data therefore requires the use of the 
formulas presented in \cite{Cordero-Carrion:2018xre,Cordero-Carrion:2019qtu}, 
which are a generalization of the well-known Casas-Ibarra parametrization 
\cite{Casas:2001sr}. 

Next we present some approximation formulas for the scalar decay
widths, since these will be useful to understand our numerical results
for the decay lengths, shown in the next section. In these estimates,
we always neglect final state masses for simplicity. Assuming $\theta
\ll 1$,\footnote{This corresponds to $\lambda_5 <1$, compare to
  eq. (\ref{eq:t2t}).}  the mass eigenstates $H^{2+}_{1,2}$ can be
identified with the gauge eigenstates to a good approximation.  The
two-body decay widths, neglecting lepton masses, are:
\begin{eqnarray}\label{eq:g2p}
\Gamma(H^{2+}_{i}) = 2 c_{2+}\frac{|h_{ee}|^2}{8 \pi}m_{H^{2+}_i} ,
\end{eqnarray}
with $c_{2+} \simeq \cos\theta$ ($\sin\theta$), if the $H^{2+}_i$
state is mostly singlet (triplet). Unless $|h_{ee}|$ is extremely
tiny, one expects that the 2-body decay of the mostly singlet state,
$S^{2+}_1$, is too fast to give a measurable decay length.

For the decays of the states $H^{3+}$ and $H^{4+}$, the decay widths
depend mainly on the mass ordering of the exotic states in our model.
If any $m_{F_i} < m_{H^{n+}}$ the decays of the scalars are 2-body
and, again, one expects them to be too fast to leave experimentally
interesting decay lengths. Similarly, if $m_{S_1} < m_{S_3}$, the
multi-charged scalars will promptly decay to $H^{2+}_1 \simeq
S^{2+}_1$ plus $W$-boson(s).  However, the situation is very different
if $m_{S_3}< m_{S_1},m_{F_1}$.  Let us estimate the widths in this
phenomenologically more interesting case.

Consider first $H^{3+}$.  For this state we need to take into account
two possible final states. The simpler one is the 3-body decay
$H^{3+}\to W^+l^+_{\alpha}l^{+}_{\beta}$. One can derive the simple
estimate:
\begin{equation}\label{eq:StoWll}
\Gamma(H^{3+}\to W^+ l^+_{\alpha}l^+_{\beta}) \simeq
\frac{g_2^2}{2}\frac{|(h_{ee})_{\alpha\beta}|^2\sin^2\theta}{f(3)}
\frac{(m_{H^{3+}}^2-m_W^2)^{7/2}}{m_{H^{2+}_1}^4m_W^2}
\end{equation}
Note the factor $1/m_W^2$, which is due to the massive gauge boson in
the final state. The term $f(3)$ corresponds to the 3-body phase space
factor for massless particles: 
\begin{equation}\label{eq:MLPS}
f(n) = 4 (4\pi)^{(2n-3)} (n-1)! (n-2)! 
\end{equation}
Equation (\ref{eq:StoWll}) represents a rough estimate for the true
value of the width in the case $m_{S_1}^2 \gg m_{S_3}^2$ and
$m_{H^{3+}} \gg m_W$.

The second important decay mode of $H^{3+}$ is the 4-body decay:
$H^{3+}\to l^+_{\alpha}l^{+}_{\beta}l^{+}_{\gamma}\nu_{\delta}$.  This
decay proceeds via diagrams involving the exotic fermions $F_k^{2+}$
and $F_k^{3+}$, such as, for example: $H^{3+}\to l^{+}_{\gamma}
(F_k^{2+})^* \to l^{+}_{\gamma}\nu_{\delta}(H^{2+}_i)^* \to
l^+_{\alpha}l^{+}_{\beta}l^{+}_{\gamma}\nu_{\delta}$.  In principle,
the width contains a double sum: Summing over the different charge
states $F_k^{2+}$ and $F_k^{3+}$, as well as over the generation
$k$. For the latter, we define an effective reduced coupling:
\begin{equation}\label{eq:effhfhfb}
|\sum_k (\frac{h_F^{\gamma k}h_{\bar F}^{\delta k}+h_F^{\delta k}h_{\bar F}^{\gamma k}}
{m_{F_k}})| \equiv |\langle\frac{h_{F{\bar F}}^{\gamma\delta}}{ m_F} \rangle|
\end{equation}
Note that in case that all $m_{F_k}\equiv m_F$, the effective $\langle
m_F \rangle$ simply reduces to $m_F$ and can be taken out from the
sum.  With eq. (\ref{eq:effhfhfb}), the 4-body decay width can be 
estimated to be very roughly:
\begin{equation}\label{eq:s3p4body}
\Gamma(H^{3+}\to l^+_{\alpha}l^{+}_{\beta}l^{+}_{\gamma}\nu_{\delta}) 
\simeq \frac{c_3}{f(4)}
|\langle\frac{h_{F{\bar F}}^{\gamma\delta}}{ m_F} \rangle|^2
\frac{|(h_{ee})_{\alpha\beta}|^2}{m_{H_2^{2+}}^4} m_{H^{3+}}^7,
\end{equation}
with\footnote{We have estimated only the contribution from one diagram
  and adjusted the prefactor $c_3$ to fit the absolute value of the
  numerical output of {\texttt Madgraph5 v2.3.3} \cite{Alwall:2011uj},
  see the next section, since eq. (\ref{eq:s3p4body}) is used only to
  understand the parameter dependence of the width.}  $c_3 \sim
(1/4)$.  Note that, there is a contribution to 4-body final states
with $\gamma=\delta$ from a diagram with $H^{3+}\to (W^+)^*
l^+_{\alpha}l^{+}_{\beta} \to
l^+_{\alpha}l^{+}_{\beta}l^{+}_{\gamma}\nu_{\gamma}$.  This diagram
can be estimated from eq. (\ref{eq:StoWll}), but replacing $f(3)\to
f(4)$ and multiplying the width by Br$(W\to l+\nu)$. The two diagrams
can interfere, thus the approximate expressions become unreliable if
both diagrams give numerical values of the same order of magnitude.

\begin{figure}
\begin{center}
\begin{tabular}{cc}
\includegraphics[width=0.27\textwidth]{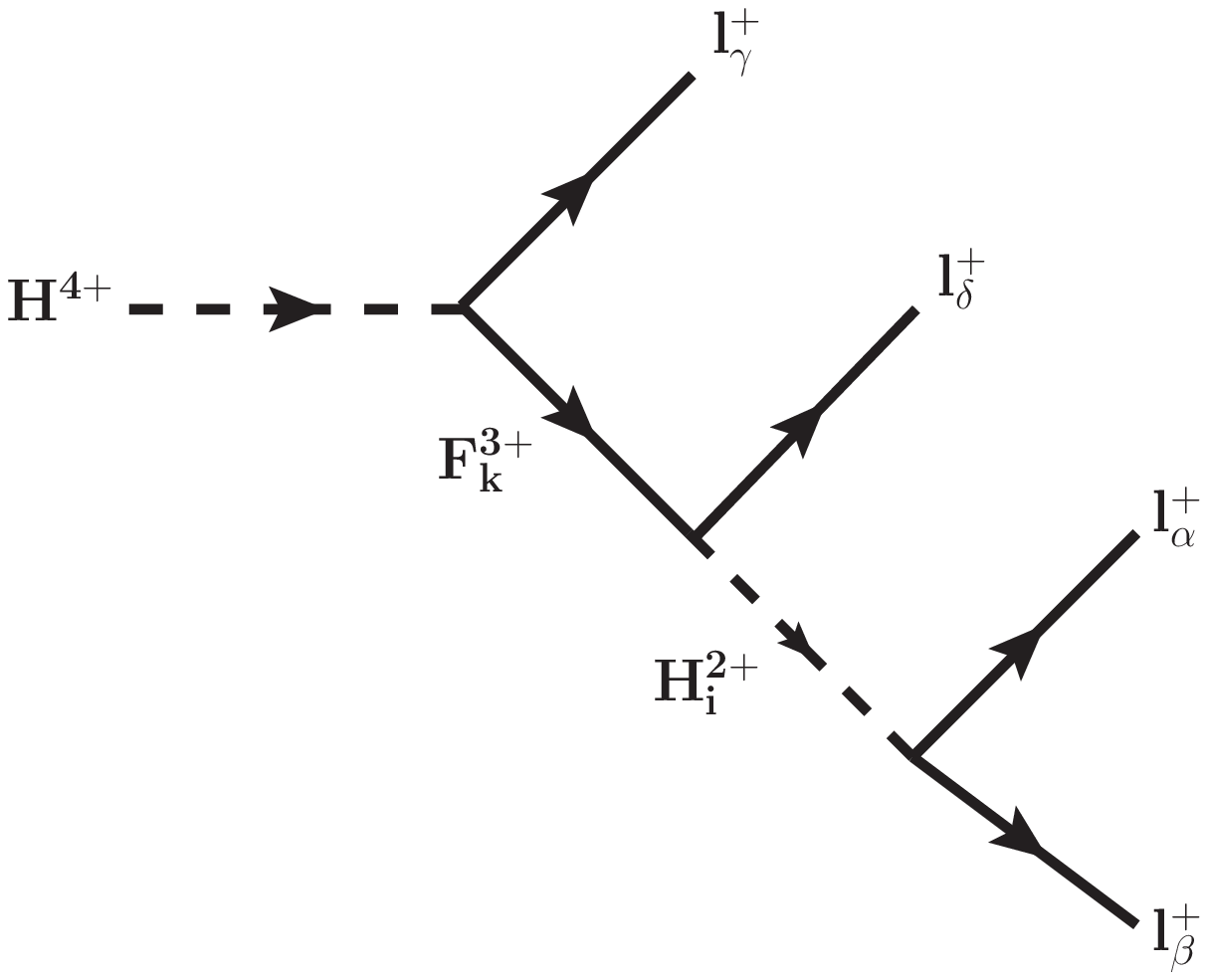}\hskip9mm
\includegraphics[width=0.27\textwidth]{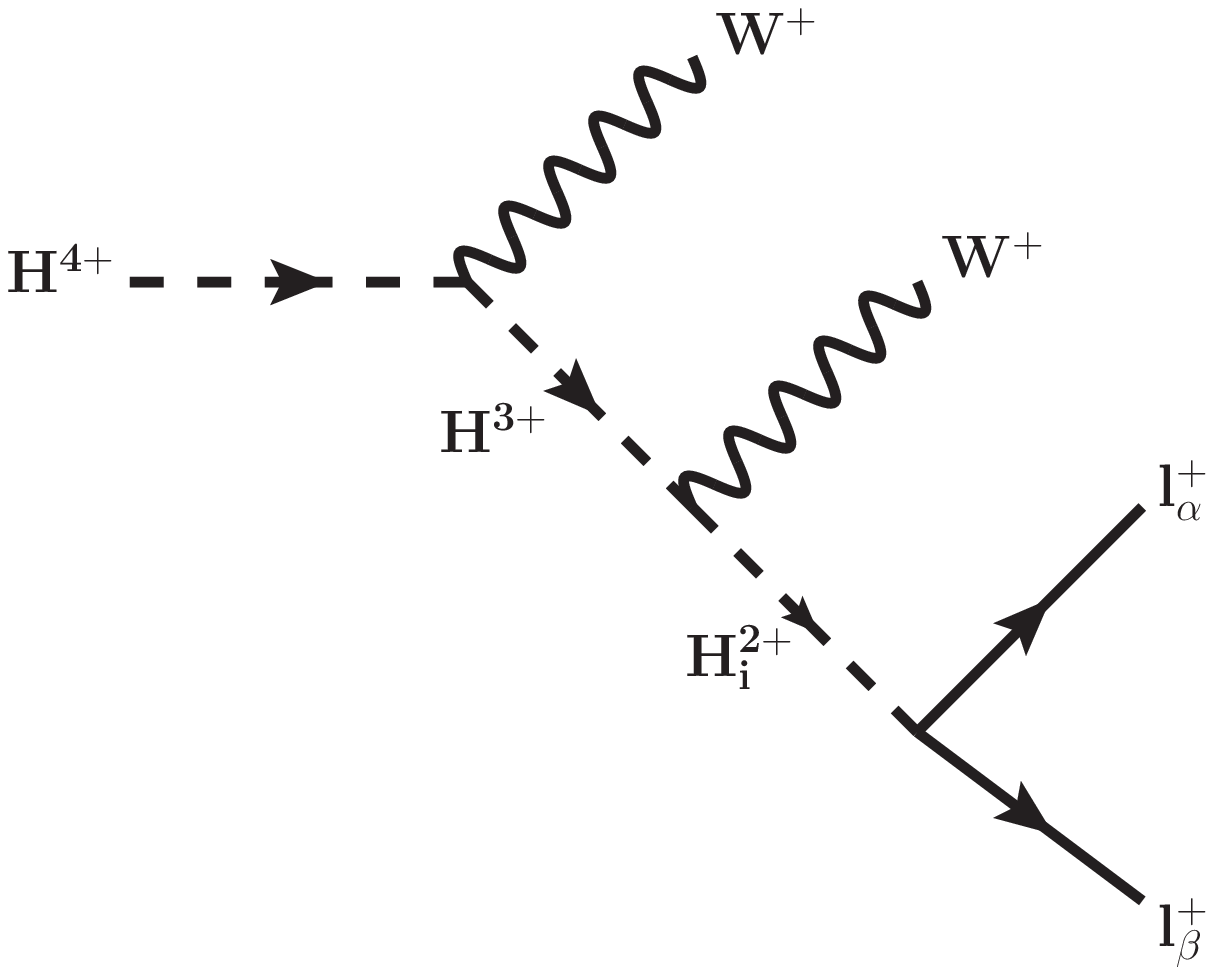}\hskip9mm
\includegraphics[width=0.27\textwidth]{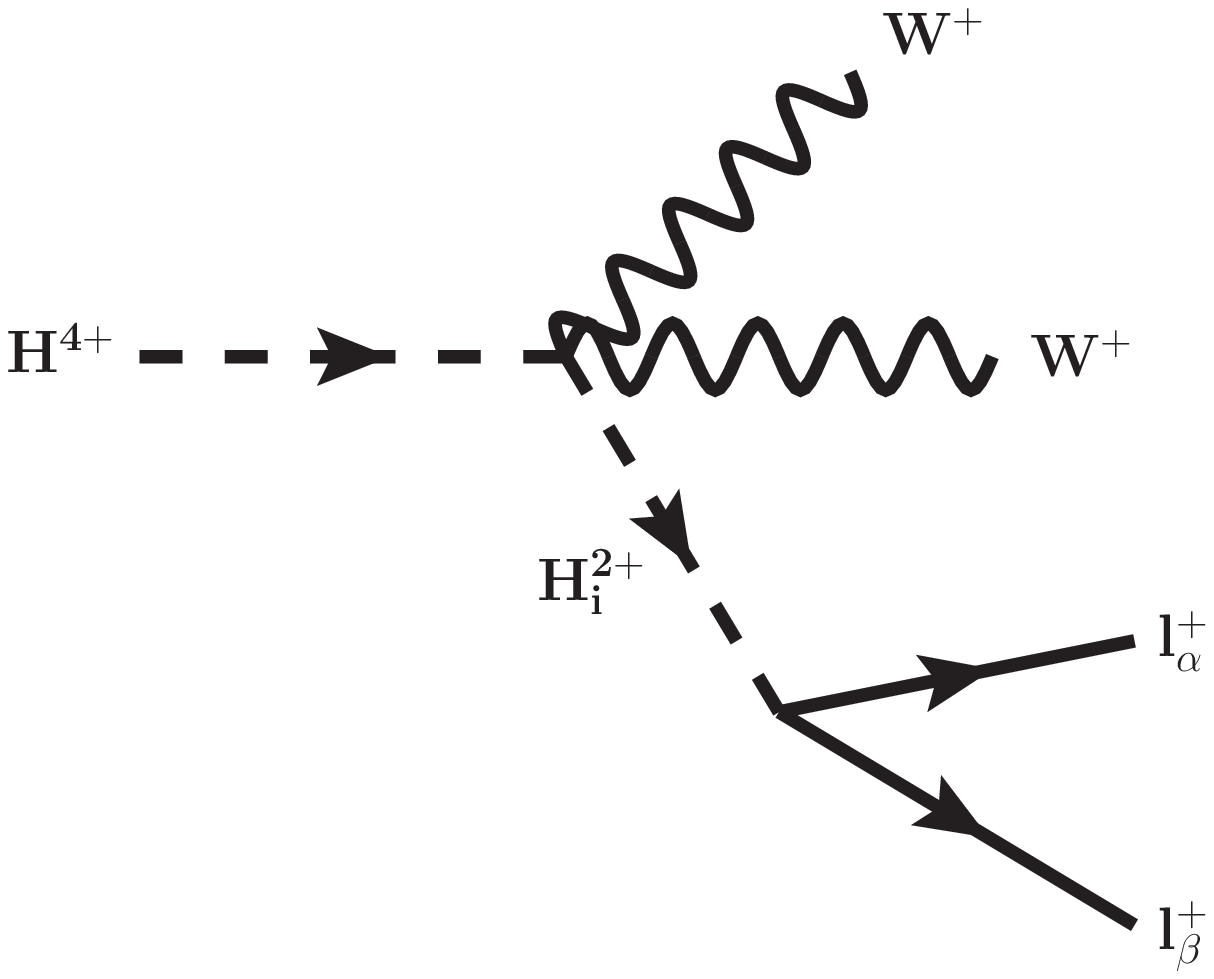}
\end{tabular}
\caption{Feynman diagrams for the decays 
$H^{4+}\to l^{+}_{\gamma}l^+_{\delta}l^+_{\alpha}l^+_{\beta}$ and
$H^{4+}\to W^+W^+l^+_{\alpha}l^+_{\beta}$. 
\label{fig:diagsS4}.}
\end{center}
\end{figure}

If $m_{S_3}< m_{S_1}$ and $m_{S_3}<m_{F_1}$, $H^{4+}$ has only 4-body
decays. There are two different final states to consider: $H^{4+}\to
l^+_{\alpha}l^{+}_{\beta}l^{+}_{\gamma}l^+_{\delta}$ and $H^{4+}\to
l^+_{\alpha}l^{+}_{\beta}W^+W^+$.  The decay to charged leptons
proceeds through similar diagrams as discussed for $H^{3+}\to
l^+_{\alpha}l^{+}_{\beta}l^{+}_{\gamma}\nu_{\delta}$ above, see also
fig. (\ref{fig:diagsS4}). The partial width is therefore estimated as:
\begin{equation}\label{eq:s4p4l}
\Gamma(H^{4+}\to l^+_{\alpha}l^{+}_{\beta}l^{+}_{\gamma}l^+_{\delta}) 
\simeq \frac{c_4}{f(4)}
|\langle\frac{h_{F{\bar F}}^{\gamma\delta}}{ m_F} \rangle|^2
\frac{|(h_{ee})_{\alpha\beta}|^2}{m_{H_2^{2+}}^4} m_{H^{4+}}^7,
\end{equation}
with $c_4 \sim (1/4)$, again simply fitted to the
numerical result.

The decay $H^{4+}\to l^+_{\alpha}l^{+}_{\beta}W^+W^+$ is caused by two
different kind of Feynman diagrams, see
fig. (\ref{fig:diagsS4}).  The vertex $HHWW \propto g_2^2$,
while $HHW \propto g_2\partial_\mu$. Assuming $H^{2+}_2 \simeq S_1$ 
is much heavier than $H^{2+}_1 \simeq S_3^{2+}$ and further 
simplifying $m_{H^{4+}} \simeq m_{H^{3+}}\simeq m_{H^{2+}_1}$, the 
middle diagram in fig. (\ref{fig:diagsS4}) is estimated 
to generate a width of the order of
\begin{equation}\label{eq:s4p4l}
\Gamma(H^{4+}\to W^+W^+l^+_{\alpha}l^{+}_{\beta})
\sim \frac{1}{f(4)}
\frac{(g_2^2\sin\theta|(h_{ee})_{\alpha\beta}|)^2}
{m_W^4} m_{H^{4+}}^5.
\end{equation}
We stress that this approximation fails rather badly if $m_{H^{4+}}
\simeq m_{H^{3+}}\simeq m_{H^{2+}_1}$ is not true.

Finally, consider the case that one of the exotic fermionic 
states is the lightest new particle: $m_{F_1} < m_{S_1},m_{S_3}$. 
Electroweak radiative corrections generate a small mass splitting 
between the $F^{3+}$ and $F^{2+}$ members of the $F_{1,2,5/2}$ 
multiplet(s). Using the results of \cite{Cirelli:2005uq,Franceschini:2008pz}, 
we estimate $\Delta M = m_{F^{3+}}- m_{F^{2+}} \sim 1.6$ GeV for 
$m_{F}\gsim 500$ GeV. The  $F^{3+}$ then has a 2-body decay 
mode $F^{3+}\to F^{2+}+\pi^+$, with a width of order ($2-3$) meV, 
again using the formulas of \cite{Cirelli:2005uq,Franceschini:2008pz}.
Thus, $F^{3+}$ decays can not be long-lived.

For $F^{2+}$, on the other hand, the main decay channel is to 
3 standard model leptons. This width is estimated as:
\begin{equation}\label{eq:Fto3f}
\Gamma(F^{2+}_1 \to l^+_{\alpha}l^+_{\beta}\nu_{\delta}) \simeq
\frac{|(h_F)_{\delta 1} (h_{ee})_{\alpha\beta}|^2}{512 \pi^3}
\frac{m_{F^{2+}_1}^5}{m_{S_1}^4}.
\end{equation}
We now turn to a discussion of the numerical results.

\section{Numerical results}
\label{sect:res}

We have implemented the model discused in section \ref{sect:model} 
in \texttt{SARAH} \cite{Staub:2012pb,Staub:2013tta}. SARAH allows 
to automatically generate
\texttt{SPheno} routines \cite{Porod:2003um,Porod:2011nf} with 
which one can do a 
numerical evaluation of mass spectra, mixing matrices, 2-body and
fermionic 3-body decays. \texttt{SARAH} also generates model
files for \texttt{MadGraph}
\cite{Alwall:2007st,Alwall:2011uj,Alwall:2014hca}. 
We use \texttt{MadGraph} to 
numerically calculate the 3-body and 4-body decay widths of the 
exotic scalars, as well as production cross sections. 

\subsection{Decay lengths}
\label{sect:len}

In this section we will discuss numerical results for the decay
lengths of the new particles in our model. Let us start by stressing
again that the mass ordering of the scalars, $H^{2+}_i$, $H^{3+}$ and
$H^{4+}$, and fermions, $F^{2+}_i$, is not fixed. If $H^{2+}_1\simeq
S^{2+}_1$ is the lightest of the exotic particles, all heavier particles
will decay fast to $H^{2+}_1+ X$, where $X$ stands symbolically for
any other particle(s), and all decays in the model are most likely
prompt. We will therefore discuss in detail only the case when 
the mostly $S^{2+}_1$ state is heavier than the other particles. 

As eqs (\ref{eq:mspp}) and (\ref{eq:msmp}) show, for moderate values
of $m_{S_3}$ and large values of $\lambda_{3a,3b}$, sizeable
splittings among the members of the triplet are possible. Once the
mass splitting is bigger than the mass of the $W$, decays such as
$H^{4+}\to H^{3+}+W^+$ will dominate the total width. These decays are
not suppressed by small neutrino masses and are therefore fast. We
thus implicitly assume in the following discussion that the
particle under consideration is the lightest member of the triplet.

Before showing the numerical results, let us briefly discuss how to
fit the experimentally observed neutrino masses in our model.
Consider the neutrino mass matrix, eq. (\ref{eq:NuMat}). In our
numerical calculations we use formulas from
\cite{Cordero-Carrion:2018xre,Cordero-Carrion:2019qtu} to parametrize
the Yukawa couplings as:
\begin{eqnarray}\label{eq:master}
h_F = \frac{1}{\sqrt{2{\cal F}}}{\cal M}^{-1/2}WT{\bar D}_{\sqrt{m}}U_{\nu}^{\dagger},
\\ \nonumber
h_{\bar F} = 
\frac{1}{\sqrt{2{\cal F}}}{\cal M}^{-1/2}W^{\star}B{\bar D}_{\sqrt{m}}U_{\nu}^{\dagger}.
\end{eqnarray}
Here, $W$ and $T$ are a unitary and an upper-triangular matrix, while
$B=(T^T)^{-1}(\id_3 + K)$, with $K$ an anti-symmetric square matrix
\cite{Cordero-Carrion:2019qtu}. ${\bar D}_{\sqrt{m}}$ and
$U_{\nu}^{\dagger}$ are the square roots of the light neutrino mass
eigenvalues and the neutrino mixing matrix, respectively. ${\cal M}$
and ${\cal F}$ are defined in eq. (\ref{eq:NuMat}).
Eq. (\ref{eq:master}) is the generalization of the well-known
Casas-Ibarra parametriziation \cite{Casas:2001sr} for the case of two
independent Yukawa matrices.  Note, that if we choose the particular
simple limit of $K=0$ and $T$ diagonal (${\hat T}$), the three vectors
in $h_F$ and $h_{\bar F}$ are fixed only up to a constant $T_{ii}$
each: $h_F \to {\hat T} h_F$ with $h_{\bar F} \to {\hat T}^{-1}
h_{\bar F}$ leaves the neutrino masses unchanged.  Essentially,
eq. (\ref{eq:master}) fixes a relation between the measured neutrino
data, see for example \cite{deSalas:2017kay}, and the Yukawa couplings
$h_F$ and $h_{\bar F}$. Once $\Delta m_{\rm Atm}^2$, $\Delta
m_{\odot}^2$ and the neutrino mixing angles are fixed from data, for
any set of scalar and fermion masses, plus a chosen neutrino mass
scale $m_{\nu_1}$, $h_F$ and $h_{\bar F}$ can be fixed as function of
$\lambda_5$.

A simple estimate for the neutrino
mass can be obtained, assuming all heavy masses approximately equal
to some scale $\Lambda$ and $\lambda_5\ll 1$. Roughly,
\begin{equation}\label{eq:EstNu}
m_{\nu} \simeq 0.05 \Big(\frac{\lambda_5}{10^{-6}} \Big)
\Big(\frac{h_F}{10^{-2}}\Big) \Big(\frac{h_{\bar F}}{10^{-2}}\Big) 
\Big(\frac{\rm 1 \hskip1mm TeV}{\Lambda}\Big) 
\hskip2mm {\rm eV}.
\end{equation}
Neutrino masses only require that the product of $\lambda_5 h_F
h_{\bar F}$ is small, but do not fix the ratios of these parameters.
Recall, $\sin 2\theta \propto \lambda_5$. We do not use eq. 
(\ref{eq:EstNu}) in our numerical studies. It serves only to show 
the order of magnitude of the parameters, necessary to explain 
neutrino data. 

Let us discuss now scalar decays.  Experimentally, the most
interesting decays are those of the $H^{4+}$. As discussed in section
\ref{sect:model}, for $H^{4+}$ we have to consider two final states:
$H^{4+}\to l^{+}_{\gamma}l^+_{\delta}l^+_{\alpha}l^+_{\beta}$ and
$H^{4+}\to W^+W^+l^+_{\alpha}l^+_{\beta}$. Pair production of
$H^{4+}H^{4-}$, see section \ref{subsect:Cross}, will then lead to
final states $(4 l^+)(4l^-)$, $(2 l^+2l^-+2W^+2W^-)$ and $(4
l^\pm)(2l^\mp)+2W^\mp$.  The last one of these is the most interesting
one, since it allows to demonstrate the existence of LNV
experimentally, for those events where both the Ws decay
hadronically.\footnote{Since Br($W\rightarrow$ hadrons)$\simeq 0.67$
  \cite{Tanabashi:2018oca}, this corresponds to about half of all
  events.  The remaining events contain leptonic decays of the $W$,
  i.e.  missing energy in the final state.}

The final state $6l+2W$ can occur with measurable rates only if
$\Gamma(H^{4+}\to 4l^+) \sim \Gamma(H^{4+}\to 2 l^+ 2 W^+)$, since
otherwise the branching ratio into either the $4l$ or $2l+2W$ final
state is likely to be too small to be observed. Recall from the
discussion in the previous section, $\Gamma(H^{4+}\to 4l^{+})\propto
h_F h_{\bar F}$, while $\Gamma(H^{4+}\to 2l^++2W^+)\propto \lambda_5$.

\begin{figure}[tbph]
\begin{centering}
\includegraphics[scale=0.5]{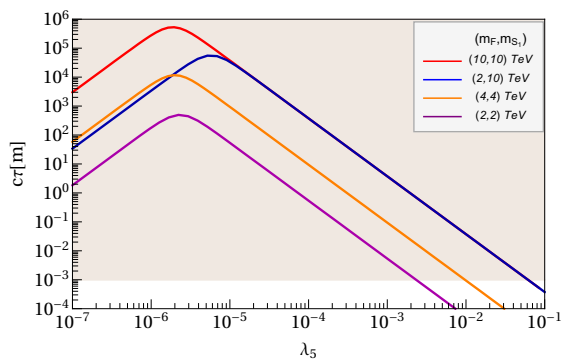}
\end{centering}
\protect\caption{\label{fig:Gam4}Decay length of $H^{4+}$ as a
  function of $\lambda_5$, for fixed choices of other parameters.  The
  lightest neutrino mass was fixed to $m_{\nu_1}=1$ meV, see also
  the text.}
\end{figure}

In fig. (\ref{fig:Gam4}) we show $c\tau$ for the decay of $H^{4+}$ as
a function of $\lambda_5$ for three sets of arbitrary masses, but other
parameters fixed. The red/blue/orange/purple curves correspond to
($m_F,m_{S_1}$) equal to ($10,10$)/($2,10$)/($4,4$)/($2,2$) TeV,
respectively.  We have also fixed $m_{H^{4+}}=1$ TeV and
$(h_{ee})_{11}=0.1$ ($\forall (h_{ee})_{ij}=0$).\footnote{This choice
  was only motivated to speed up the numerical calculations. The total
  width always scale perfectly with $|h_{ee}|=\sum
  \sqrt{(h_{ee})_{ij}^2}$.  }  
The Yukawa couplings $h_F$ and $h_{\bar F}$ are calculated in each
point using eq. (\ref{eq:master}) to explain the best fit point of
neutrino oscillation data. The plot corresponds to $m_{\nu_1} =
10^{-3}$ eV (and normal hierarchy). There is only a moderate
dependence of the decay lengths on the choice of the lightest neutrino
mass. As the plot shows, the decays of the $H^{4+}$ are slow and long
charged tracks are to be expected in large parts of the parameter
space, unless $\lambda_5 \to 1$ or $\lambda_5 \to 0$.

The plots in fig. (\ref{fig:Gam4}) show that $c\tau$ is maximized for 
intermediate values of $\lambda_5$. This is easily understood from 
our neutrino fit. Consider eq. (\ref{eq:EstNu}). Keeping the neutrino 
mass fixed, lower values of $\lambda_5$ require an increase of 
$h_Fh_{\bar F}$. This increases $\Gamma(H^{4+}\to 4l^{+})$. Equally, 
large $\lambda_5$ increases $\Gamma(H^{4+}\to 2l^++2W^+)$. The maximal 
$c\tau$ therefore corresponds to the point where both partial 
widths are equal, and minimized in absolute value. In other words, 
the possibility to observe LNV experimentally is largest for the 
points with the largest $c\tau$. 

One word of caution. One should not take the decay lengths in 
fig. (\ref{fig:Gam4}) as fixed predictions, since $(h_{ee})$ is 
not fixed by neutrino data. Since all decay widths of $H^{4+}$ 
are proportional to $(h_{ee})^2$ both, larger and smaller, $c\tau$ 
than the ones shown in  (\ref{fig:Gam4}) are allowed from the 
model. Note, however, that $(h_{ee})$ can not be arbitrarily 
small, since big bang nucleosynthesis disfavours particles 
with life-times larger than ${\cal O}(0.1-1)$ s
\cite{Aghanim:2018eyx,Kawasaki:2004qu,Jedamzik:2006xz}.

\begin{figure}[tbph]
\begin{centering}
\begin{tabular}{c c}
\includegraphics[scale=0.42]{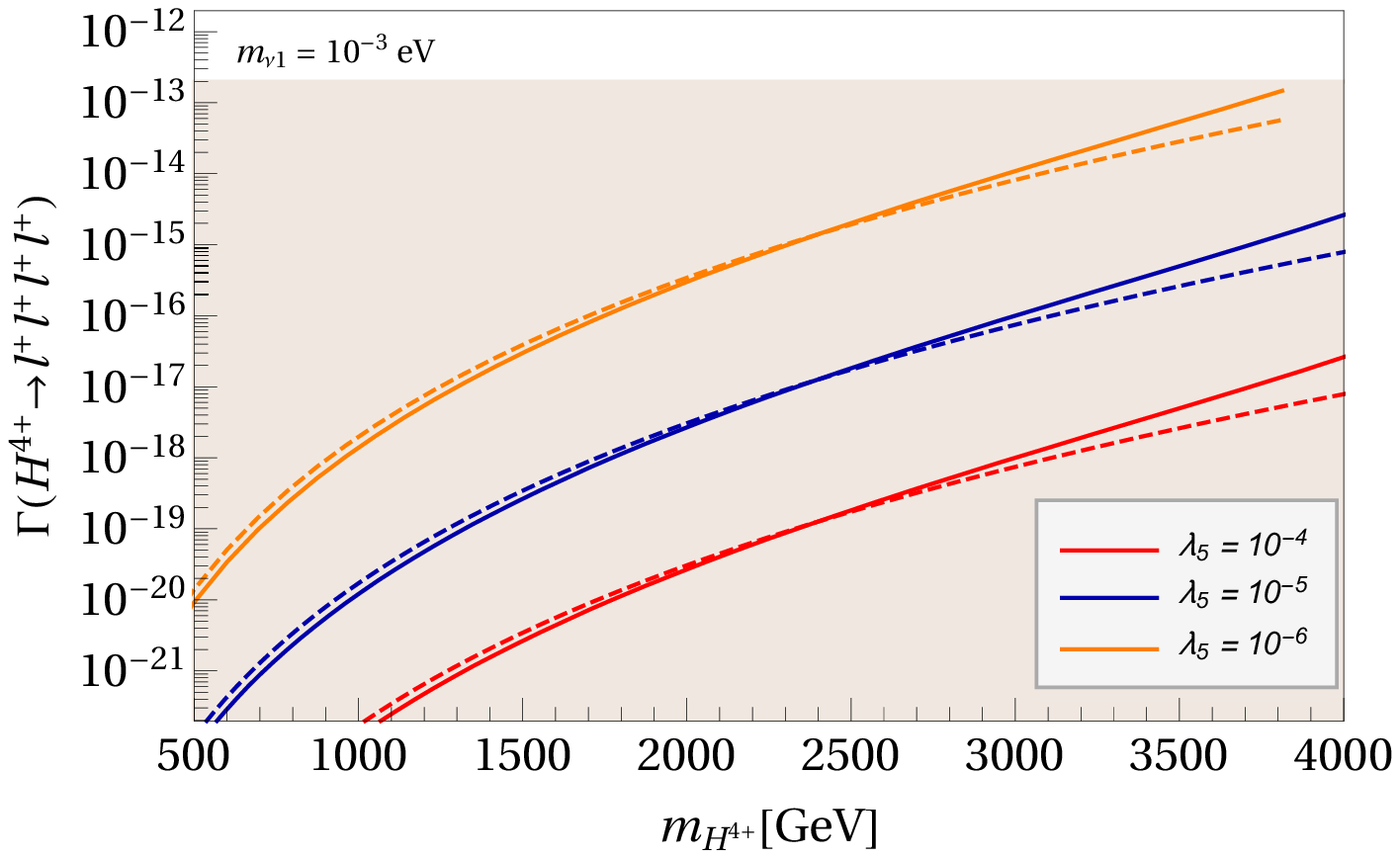}&
\includegraphics[scale=0.42]{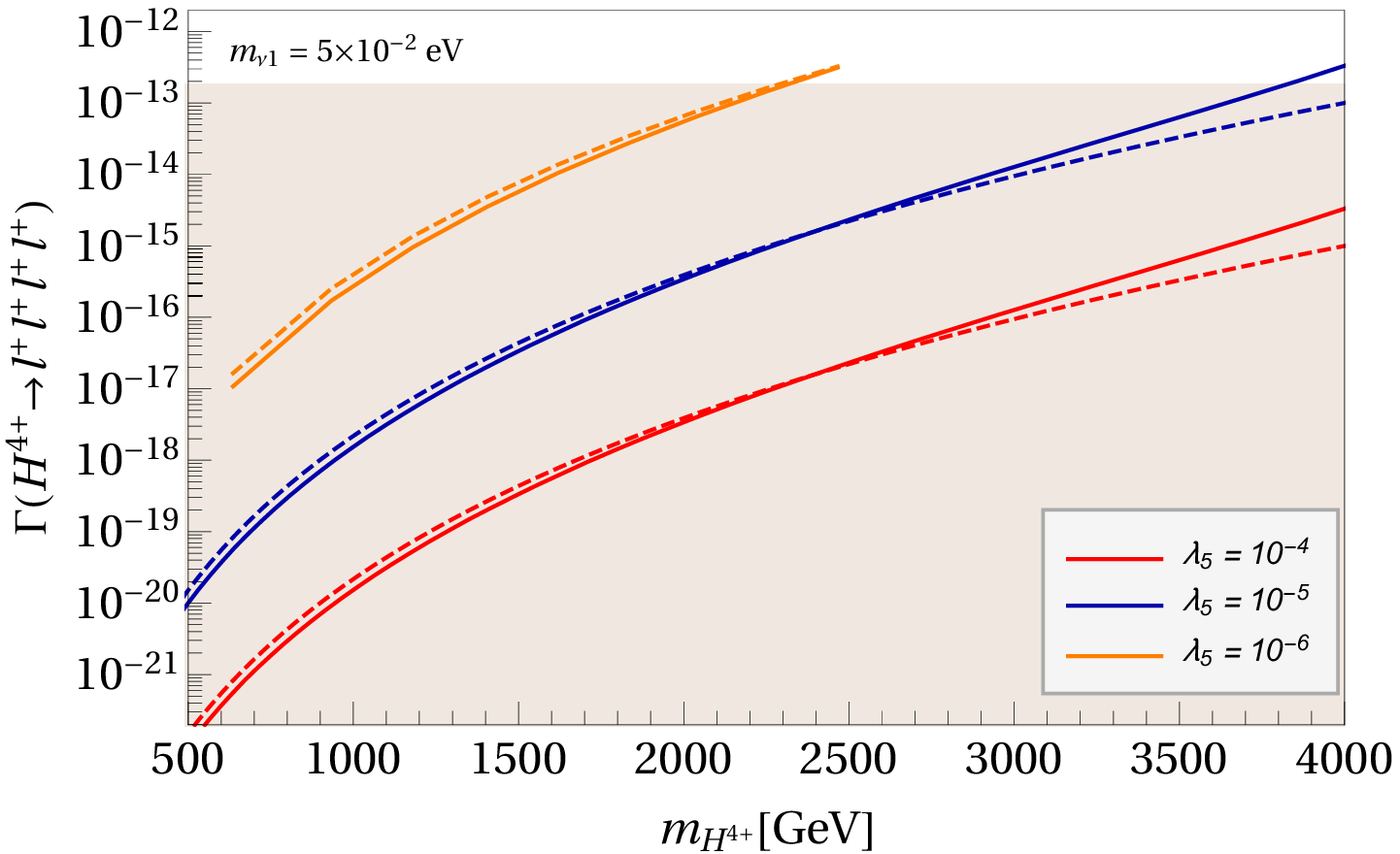}\\
\end{tabular}
\end{centering}
\protect\caption{\label{fig:Gam4l}Decay width of $H^{4+}\to 4l$ as a
  function of $m_{H_{4}}$, for fixed choices of other parameters.
  $(h_{ee})_{11}= 1$, $m_F$ = 5 TeV, $m_{S_1}=5$ TeV and $\lambda_5 =
  10^{-4},10^{-6},10^{-8}$ (red,blue,orange). $h_{FFB}$ chosen to fit
  the neutrino masses. Solid and dashed lines correspond to numerical calculations and analytical estimations respectively.}
\end{figure}

\begin{figure}[tbph]
\begin{centering}
\begin{tabular}{c c}
\includegraphics[scale=0.42]{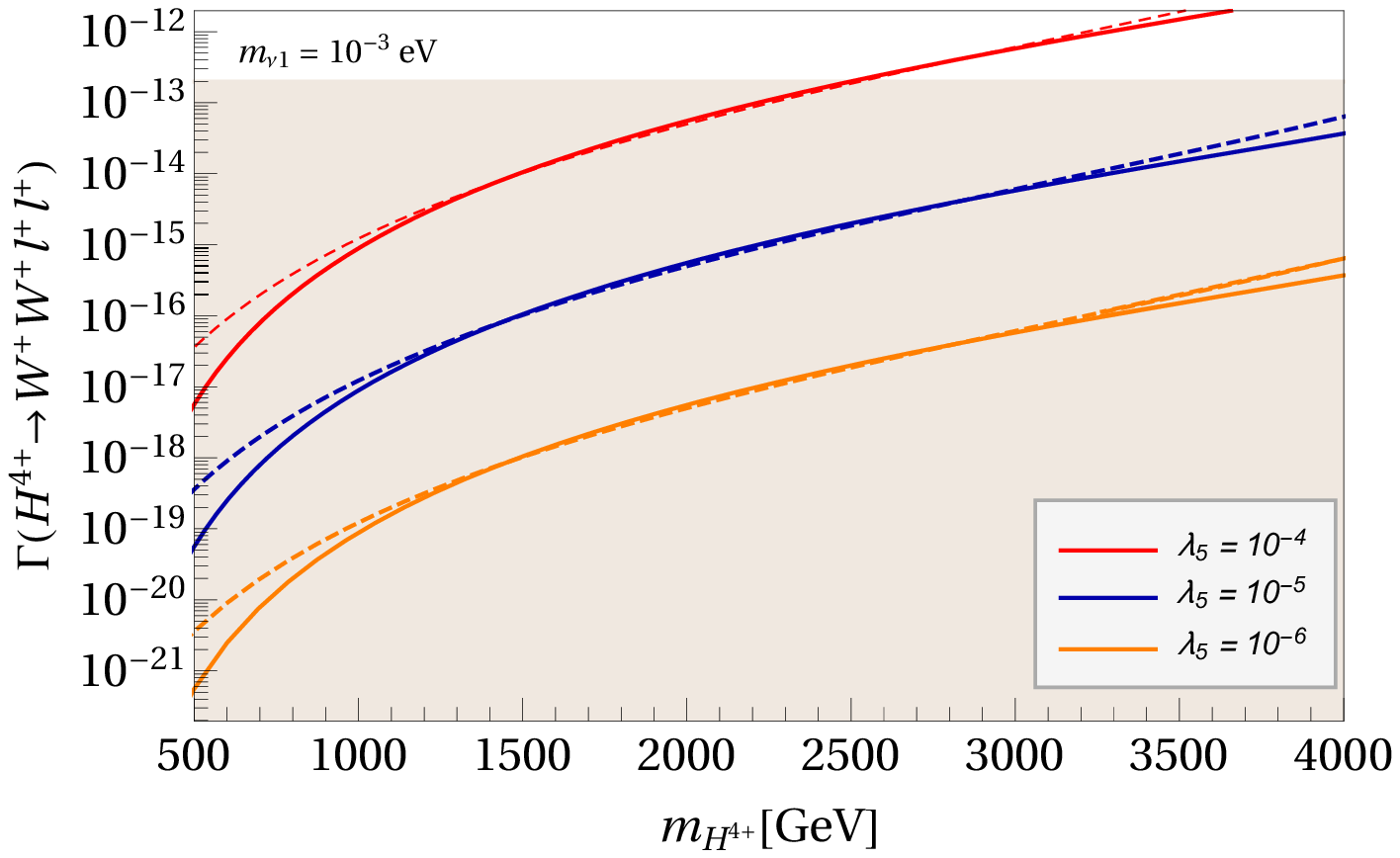}&
\includegraphics[scale=0.42]{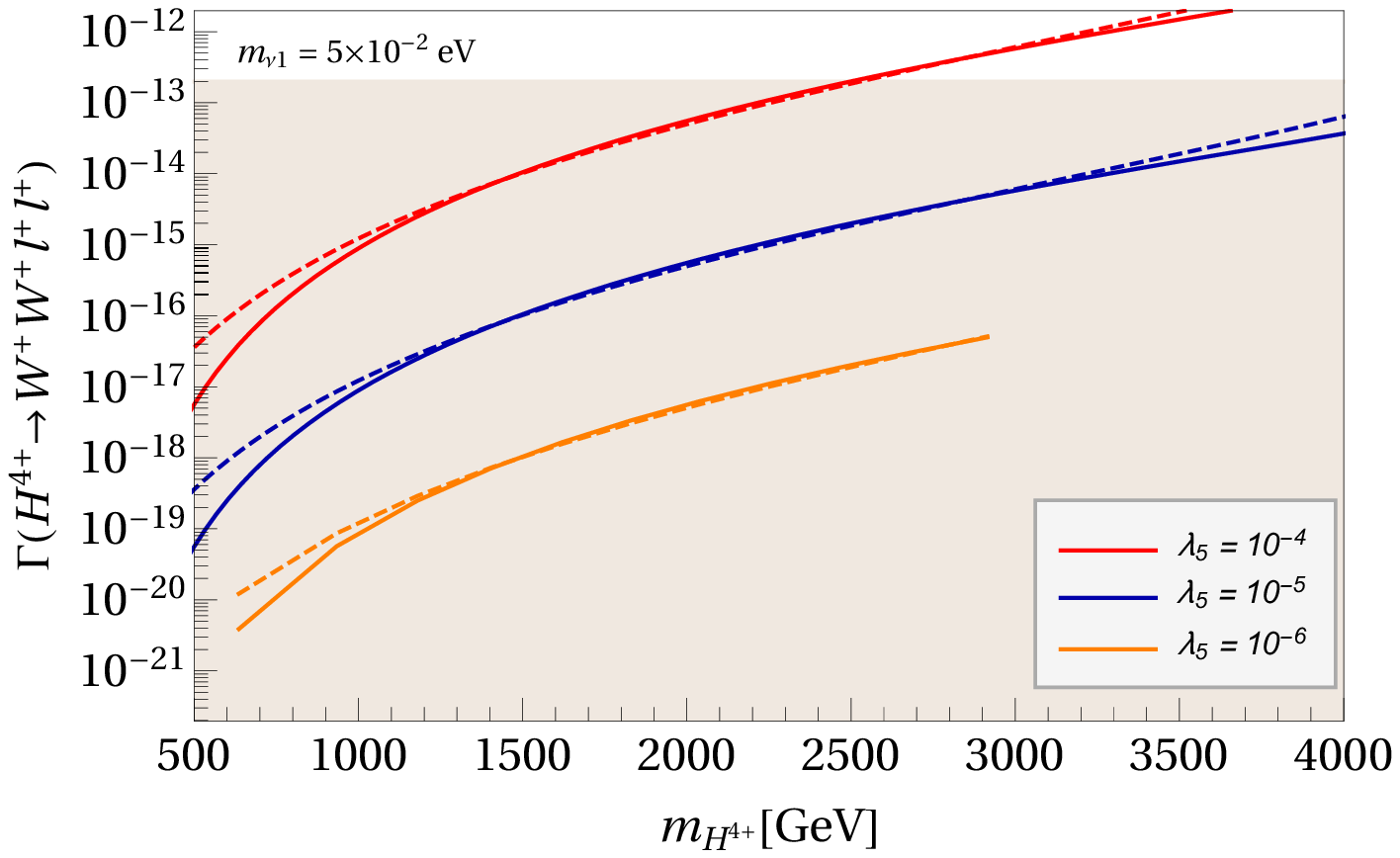}
\end{tabular}
\end{centering}
\protect\caption{\label{fig:Gam2w2l}Partial decay width of $H^{4+}\to
  2W2l$ as a function of $\lambda_5$, for fixed choices of other
  parameters. Parameters as in fig. (\ref{fig:Gam4l}).}
\end{figure}

Figs (\ref{fig:Gam4l}) and (\ref{fig:Gam2w2l}) show the partial widths
for $H^{4+}$ for the final states $H^{4+}\to 4l$ and $H^{4+}\to 2W2l$,
respectively, as a function of $m_{H^{4+}}$ for fixed choice of other
parameters.  The plots serve only to demonstrate the strong dependence
of the widths on $m_{H^{4+}}$. However, measurably small widths can
occur even for the largest masses shown. Note, that the plots show
partial widths and not decay lengths.  The plots also demonstrate that
there is some dependence of the widths on the absolute scale of
neutrino mass.  Since the lightest neutrino mass is currently unknown
(only upper limits exist), larger lengths than shown in the plot are
possible.  (Again, the same is true for $h_{ee}$, see discussion
above.)

\begin{figure}[tbph]
\begin{centering}
\includegraphics[scale=0.5]{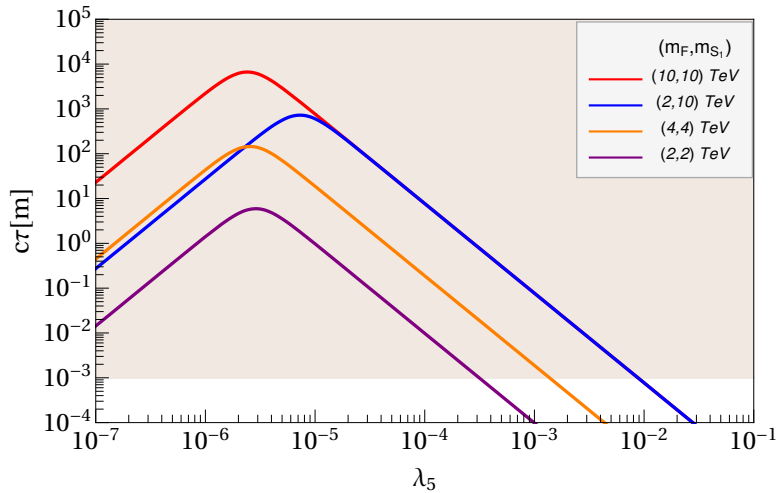}
\end{centering}
\protect\caption{\label{fig:Gam3}Decay length of $H^{3+}$ as a 
function of $\lambda_5$, for fixed choices of other parameters. 
All other parameters as in fig. (\ref{fig:Gam4}).}
\end{figure}

We now turn to the discussion of the decay lenghs for $H^{3+}$.
Fig. (\ref{fig:Gam3}) shows $c\tau$ for $H^{3+}$ as a function of
$\lambda_5$ for the same parameters as used in fig. (\ref{fig:Gam4}).
In the part of parameter space where the 4-body final state $H^{3+}\to
3l+\nu$ dominates, the decay lengths of $H^{3+}$ are simular to those
of $H^{4+}$, whereas for larger values of $\lambda_5$, when $H^{3+}\to
W+2l$ dominates, the lengths of $H^{3+}$ are expected to be around two
orders of magnitude smaller than those for $H^{4+}$. The overall
dependence of parameters is, however, very similar to those discussed
for $H^{4+}$. Note, however, that the LNV final state for
$H^{3+}H^{3-}$ pair production involves always at least one neutrino,
and thus can not be used to demonstrate LNV experimentally.

Finally, let us briefly discuss the decays $H^{2+}_1 \to
l^+_{\alpha}l^+_{\beta}$. Since these are 2-body decays, the $c\tau$ of 
$H^{2+}_1$ is always shorter than those found for $H^{3+}$ and $H^{4+}$. 
Nevertheless, for the $H^{2+}_1 \simeq S^{2+}_3$ state, one can 
estimate
\begin{equation}\label{ctauH2}
c\tau(H^{2+}_1) \simeq 1 \Big(\frac{0.1}{|h_{ee}|}\Big)^2 
 \Big(\frac{10^{-5}}{|\lambda_5|}\Big)^2 
\Big(\frac{1 \hskip1mm{\rm TeV}}{m_{H_1^{2+}}}\Big)
\Big(\frac{m_{S_1}}{2 \hskip1mm{\rm TeV}}\Big)^4 \hskip1mm {\rm cm}.
\end{equation}

Before closing this subsection, we mention that one expects that 
the fermion decays are rather fast. Let us assume for this estimate 
that one of the fermions is the lightest exotic particle. One then 
can make a rough guess that
\begin{equation}\label{ctauF}
c\tau(F_1) \simeq 5 \times 10^{-8} \Big(\frac{0.1}{|h_{ee}|}\Big)^2 
 \Big(\frac{10^{-2}}{|h_F|}\Big)^2 
\Big(\frac{1 \hskip1mm{\rm TeV}}{m_{F_1}}\Big)
\Big(\frac{m_{S_1}}{2 \hskip1mm{\rm TeV}}\Big)^4 \hskip1mm {\rm m}.
\end{equation}
Thus, unless $|h_F|$ is very small or $m_{S_1}$ very large, one 
expects that the exotic fermions of this model decay promptly.

\subsection{Cross sections, current limits and future expectations}
\label{subsect:Cross}

In this subsection we will show the results for pair-production cross
sections for the exotic scalars of our neutrino mass model. We will
then discuss various LHC searches, from which limits on the parameter
space of the model can be currently derived. We will use these limits
to estimate the discovery prospects for the high-luminosity LHC,
i.e. for the expected future ${\cal L}=3/$ab of statistics.  We also
mention briefly the expectations for a hypothetical 100 TeV
proton-proton collider.

\begin{figure}[tbph]
\begin{centering}
\includegraphics[scale=0.65]{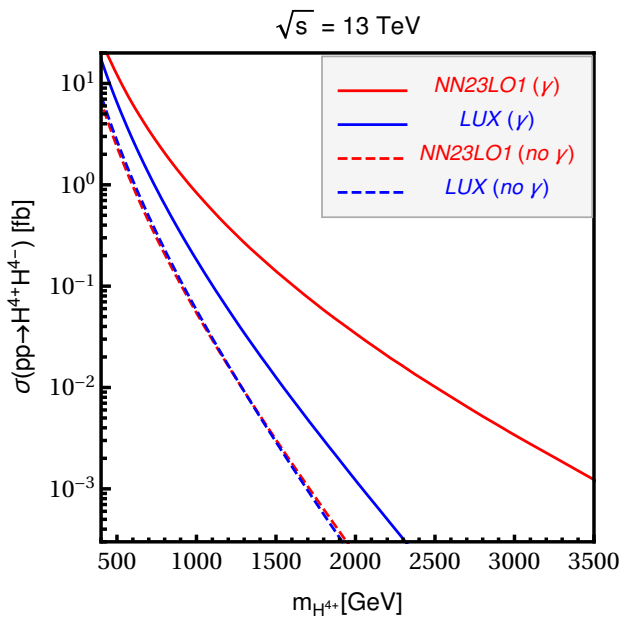}
\end{centering}
\protect\caption{\label{fig:Cxsect}Pair production cross 
sections for $H^{4+}H^{4-}$ at the LHC for $\sqrt{s}=13$ TeV using different 
PDFs. For a discussion see text.}
\end{figure}

Pair production of multiply charged particles is dominated by
photon-photon fusion diagrams, which are very important in particular
at large scalar masses, despite the tiny parton density of the photon
inside the proton. This has been discussed recently for example in
\cite{Ghosh:2017jbw}. \texttt{MadGraph} \cite{Alwall:2014hca} uses
NNPDF23LO \cite{Ball:2013hta} as the standard choice for the parton
distribution function (PDF). However, as discussed in the context of a
cross section calculation \cite{Ghosh:2018drw} for the ``BNT''
neutrino mass model \cite{Babu:2009aq}, large uncertainties in the
photon density in NNPDF23LO lead to large uncertainties in predicted
cross sections. On the other hand, Manohar et
al. \cite{Manohar:2016nzj,Manohar:2017eqh} have discussed a
model-independent determination of the photon PDF inside the proton,
leading to much smaller errors in the determination of the photon density
within the proton. The resulting {\texttt
  LUXqed17$\_$plus$\_$PDF4LHC15$\_$nnlo$\_$100} combines QCD partons
from {\texttt PDF4LHC15} \cite{Butterworth:2015oua} with the LUXqed
calculation of the photon density.

We have therefore calculated cross sections for pair production of
$H^{4+}H^{4-}$ for both, NNPDF23LO and LUXqed PDFs.
Fig. (\ref{fig:Cxsect}) shows the results for NNPDF23LO (in red) and
LUXqed17 (blue) including the photon content (full lines) and without
photon-photon fusion diagrams (dashed lines). Putting the photon
density in the proton artificially to zero, NNPDF23LO and LUXqed17
lead to very similar results, as can be seen from the figure. On the
other hand, the figure shows that NNPDF23LO leads to significantly
larger cross sections than LUXqed17, once the photon-fusion diagrams
are included in the calculation.  The plot also demonstrates that even
for the LUXqed17, with much smaller photon densities, photon-photon
fusion diagrams dominate the cross section, especially at large 
masses.

\begin{figure}[tbph]
\begin{centering}
\includegraphics[scale=0.65]{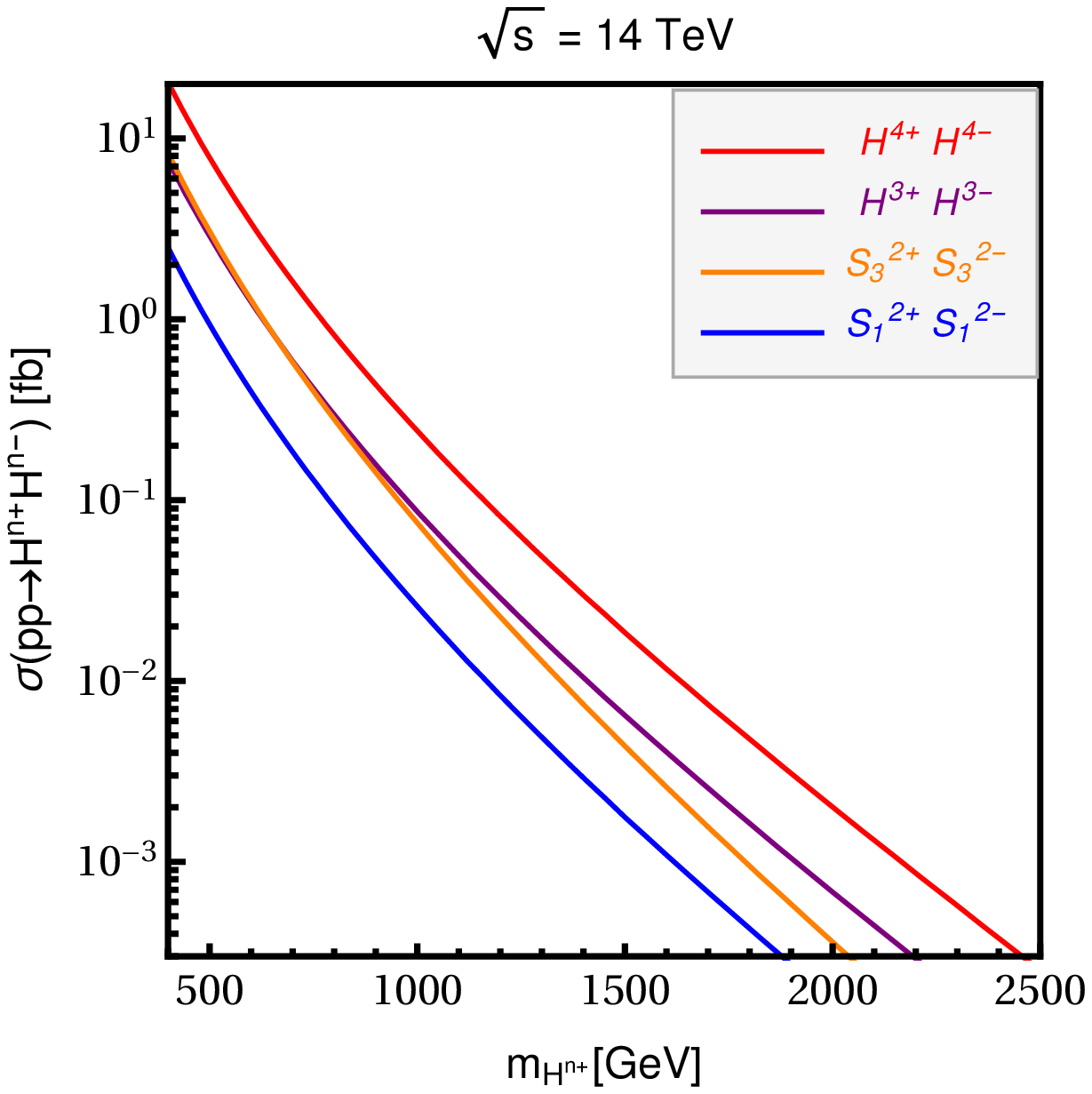}\hskip5mm
\includegraphics[scale=0.65]{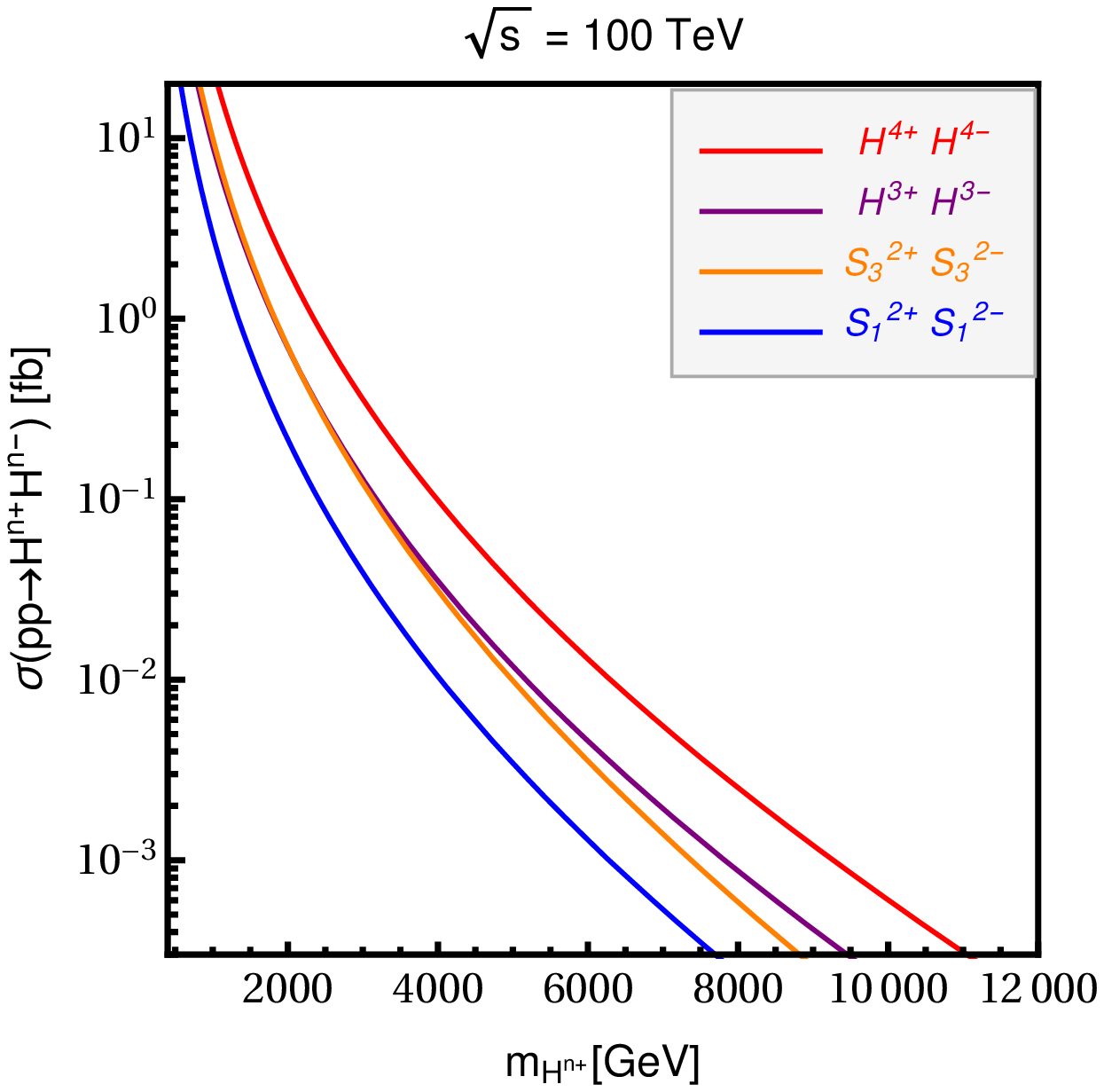}
\end{centering}
\protect\caption{\label{fig:xsect}Pair production cross sections for
  the different charged scalars in our model. The calculation was done
  with {\texttt LUXqed17$\_$plus$\_$PDF4LHC15$\_$nnlo$\_$100}
  \cite{Manohar:2016nzj,Manohar:2017eqh} and includes photon-photon
  fusion diagrams. To the left $\sqrt{s}=14$ TeV, to the right
  $\sqrt{s}=100$ TeV. }
\end{figure}

Given the importance of photon-fusion diagrams in our calculation and
considering that the LUXqed PDFs have been calculated especifically to
reduce the error in the photon density
\cite{Manohar:2016nzj,Manohar:2017eqh}, we will base our discussion on
the LHC sensitivity on cross section calculations using the {\texttt
  LUXqed17$\_$plus$\_$PDF4LHC15$\_$nnlo$\_$100} PDFs.
Fig. (\ref{fig:xsect}) shows cross sections for the four different
charged scalars as a function of their mass. The left of the figure
shows the results for $\sqrt{s}=14$ TeV, while the plot on the right
is for pp-collisions at $\sqrt{s}=100$ TeV.  We mention that, similar
to fig. (\ref{fig:Cxsect}), the cross sections calculated with the
NNPDF23LO PDF are much larger than those shown in the figure. However,
we will not use the NNPDF23LO PDF results in the following and thus,
do not show the corresponding plots.

We now turn to a discussion of LHC searches. While none of the
existing searches of ATLAS and CMS covers exactly the multi-lepton
signals we are interested in, several of the ``exotics''
searches can be used to put limits on the parameter space of our
model. We can roughly divide these searches into three categories: (i)
quasi-stable particles, i.e. $c\tau$ of the order or larger than
typical detector sizes; (ii) finite $c\tau$, say 
 $c\tau \sim {\cal O}(10^{-3}-1)$ m. And, finally, (iii) prompt 
decays.

ATLAS has searched for stable (or {\it quasi-stable}) multi-charged
particles (``MCPs'') in run-II with $\sqrt{s}=13$ TeV and ${\cal L} =
36.1$/fb of statistics \cite{Aaboud:2018kbe}.\footnote{Also CMS
  searched for long-lived charged particles
  \cite{Khachatryan:2016sfv}.  However, the CMS publication is based
  on only ${\cal L}=2.6$/fb and thus currently is not competitive with
  the ATLAS limits discussed here.}  In this search
\cite{Aaboud:2018kbe} the MCPs are assumed to live long enough to
traverse the entire ATLAS detector without decaying, and thus the
analysis exploits their muon-like signature. The search is based on
the anomalously large ionization loss of particles with $|q|=z e$ and
$z \ge 2$ that are long-lived enough to reach (and traverse) the muon
spectrometer.  This implies that the maximum efficiency of this search
is reached for $c\tau$ larger or equal roughly order $c\tau \sim 10$
m.  There is a trigger on particles with $\beta >0.6$, due to a timing
window. This requirement reduces the trigger efficiency for the
largest masses and charges. ATLAS therefore applied a second possible
trigger in this search based on $E_T^{miss}$, with $E_T^{miss} > 70,90
$ and $110$ GeV, for different data subsets.  This second trigger,
however, is responsible only for about 20 \% of the overall signal
events.

This search does not cover our multi-lepton signal, and in particular
can not establish LNV.  Nevertheless, this search can be recast to
give an estimate of the LHC sensitivity for the model parameter space,
where $c\tau \sim 10$ m or larger.  The ATLAS simulation shows that
this search is essentially background free, with an estimated number
of background events less than 0.26 [$5.1 \times 10^{-2}$] for $z=2$
[$z>2$].  ATLAS then shows, see fig. (8) of \cite{Aaboud:2018kbe},
upper limits on the production cross section as function of mass
for different choices of $z$. Experimental limits are in the range of
$(0.2-0.4)$ fb in the range of masses ($200-1400$) GeV, for $z=(2-5)$.
From our calculated cross sections we can then estimate lower 
limits on the scalar masses: $m_{H^{4+}}\gsim 980$ GeV,
$m_{H^{3+}}\gsim 820$ GeV, $m_{H^{2+}}\gsim 800$ GeV (for a mostly 
triplet state) and  $m_{H^{2+}}\gsim 660$ GeV (for a $SU(2)_L$ singlet 
state). We mention again that these limits apply only for particles 
with $c\tau$ larger than a few m.

We also want to stress the importance of the photon-fusion diagrams.
The limit on $m_{H^{4+}}$, for example, would change to
$m_{H^{4+}}\gsim 820$ GeV, excluding these diagrams from the
calculation.  Even more important is the correct choice of the
PDF. Using NNPDF23LO one would have estimated an unrealistically 
large limit of $m_{H^{4+}}\gsim 1380$ GeV.

Since this search is background free, we can make estimates for future
sensitivities by simply scaling existing limits with the expected
luminosity gain. For ${\cal L}=3/$ab, we obtain with this procedure
$m_{H^{4+}}\gsim 2$ TeV, $m_{H^{3+}}\gsim 1.7$ TeV, $m_{H^{2+}}\gsim
1.6$ TeV (triplet) and $m_{H^{2+}}\gsim 1.4$ TeV (singlet). For
completeness we mention the corresponding numbers for a $\sqrt{s}=100$
TeV collider with ${\cal L}=30/$ab: $m_{H^{4+}}\gsim 11.4$ TeV,
$m_{H^{3+}}\gsim 10$ TeV, $m_{H^{2+}}\gsim 9$ TeV (triplet) and
$m_{H^{2+}}\gsim 8$ TeV (singlet).

A search for decays of stopped exotic long-lived particles has been
published by CMS \cite{Sirunyan:2017sbs}. The search is based on
stopped particles decaying to hadronic final states in the calorimeter
or to final states involving muons in the muon spectrometer.
Statistics is ${\cal L}=38.6$/fb ($39.0$/fb) for calorimeter (MS)
search. Cross-section limits assume specific final states, non of
which is our multi-lepton signal.  Thus, these limits do not apply
directly to our model. However, we mention that the best limits are of
the order of $(20-500)$ fb for a time window of $(10^{-6}-10^5)$ sec and
it can be expected that a search for the more complicated final
states, as we are interested in, would give similar results.  At least
currently, however, these numbers are not competitive with the charged
track search discussed above.

For $c\tau$ in the intermediate region, say ($c\tau=1$ mm - 1 m),
there is currently no search in ATLAS or CMS that can be directly
converted into limits on the parameters of our model. In this $c\tau$
region, the model predicts as signal tracks with anomalously large
ionization loss, corresponding to the multiple electric charges, with
multiple leptons (with or without jets from the gauge boson decays) at
the end of the tracks.

CMS has searched for events with oppositely charged, displaced
electrons and muons at 8 TeV~\cite{Khachatryan:2014mea} (with 19.7
fb$^{-1}$) and 13 TeV~\cite{CMS:2016isf} (with 2.6 fb$^{-1}$).  Also
ATLAS has looked for displaced vertices made from pairs of leptons
decaying inside the tracker at 8 TeV (with 20.3 fb$^{-1}$)
\cite{Aad:2015rba}.  Searches for displaced lepton pairs or displaced
secondary vertices (where the leptons are associated back to the same
vertex) can be applied to derive limits on $S^{2+}$. The search by CMS
\cite{CMS:2016isf} gives currently the strongest limit. Their search
was motviated by a supersymmetric model and their lower limit on the
scalar top mass of $m_{\tilde t_1}\gsim 870$ GeV corresponds to an
upper limit of roughly $\sigma \times Br \simeq 10$ fb. Assuming 
that a same-sign lepton search would have a similar limit,
\footnote{Same-sign di-lepton searches usually have less background
  than opposite-sign di-leptons. This assumption should therefore by
  conservative.} rough estimates of $m_{H^{2+}}\gsim 370$ GeV ($270$
GeV) for mostly triplet (singlet) for a branching ratio to
electron-muon final state equal to one can be estimated from
this. Again, the limit is not very strong since the statistics is
based on only $2.6/$fb.

Both, CMS \cite{Sirunyan:2018ldc} and ATLAS \cite{Aaboud:2017mpt} have
searched for ``disappearing tracks''. These events are defined to
contain an isolated (charged) short track with\footnote{A short track
  or ``tracklet'' requires no associated hits beyond the pixel layers
  as a ``disappearance condition''.}, (i) little or no energy in
associated calorimeter deposits, and (ii) no associated hits in SCT
and muon detectors. In addition, events are selected by requiring (iii)
large $E_{miss}$ in the event, which is ensured by the presence of a
high $p_{T}$ ISR jet. While a charged track search would cover the
intermediate $c\tau$ region of our multi-lepton signature in
principle, the requirements (i)-(ii)-(iii) make it difficult that
these searches are used as constraint. Nevertheless, disappearing
track searches could have some sensitivity to ``kinked track'' signals,
as shown for example in \cite{Evans:2016zau}. Specially stronger
limits could be obtained with some trigger modifications, by cutting
directly on the measured track momentum \cite{Mahbubani:2017gjh}. This
encourages the experimental collaborations to study extensions and/or
modifications of existing disappearing track search strategies.  We
note that these searches \cite{Sirunyan:2018ldc,Aaboud:2017mpt} are
actually not background free. Adding (displaced) charged leptons (and
jets) to the track search should actually yield lower backgrounds and
we expect that limits from a dedicated search would be close to
background free and thus yield much stronger limits than those derived
in \cite{Sirunyan:2018ldc,Aaboud:2017mpt}.

Searches for multitrack displaced vertices
\cite{Aad:2015rba,Aaboud:2017iio} could, in principle, also be
sensitive to the model parameters, when the charged $H^{n+}$ decay
within the inner tracker of ATLAS and/or CMS (or within roughly
$c\tau=4-300$ [mm]). Ref. \cite{Aaboud:2017iio}, based on $32.8/$fb, gives
the strongest upper limit at $c\tau=3$ cm of $0.15$ fb. This limit,
however, is a combination of different final state searches, at least
one class of events including a high $E_{miss}$ cut. We therefore can
not directly convert this limit into a lower limit on the scalar
masses. However, it was shown that displaced vertex searches can be reinterpreted 
in the context of neutrino mass models by requiring a lepton 
trigger \cite{Cottin:2018kmq,Cottin:2018nms}. The ATLAS experiment 
performed such search when two lepton-tracks are coming from the displaced 
vertex~\cite{Aad:2019kiz}. When no other associated prompt 
objects are available to trigger on, one can require the lepton to be
explicitly associated to the displaced vertex \cite{Chiang:2019ajm}.

For the parameter region of our model where the scalars decay
promptly, various ``seesaw-searches'' can give limits on our model
parameters.  CMS has updated their seesaw type-III search in
\cite{Sirunyan:2019bgz}. This paper is based on ${\cal L}=137$/fb.
The signal in this search is {\it at least three charged leptons plus
  missing energy}. These limits therefore can be converted into limits
only for $H^{3+}$ and for that part of parameter space where $H^{4+}$
decays to the $2W2l$ final state (with the $W$'s decaying
leptonically). Limits are given unfortunately only for the
``flavour-democratic'' scenario, i.e. the decays of the heavy leptons
(charged and neutral) are assumed to have equal branching ratios for
$e$, $\mu$ and $\tau$.  Note that this limit is rather sensitive to
this assumption.  Recall, that in the earlier analysis of
\cite{Sirunyan:2017qkz} limits in the range (390-930) GeV were found,
when varying the branching ratios to the different charged lepton
generations from zero to one. The upper limit depends on the assumed
mass of the heavy state searched for and range from order $10$ fb at
500 GeV, to as low as ($2-3$) fb for masses in the range $1-1.4$ TeV.
This does not yet give meaningful limits, but will be interesting 
to constrain our exotic scalars, once more statistics is accumulated.

Our $H^{2+}$'s are also constrained by seesaw type-II searches, such
as \cite{Aaboud:2017qph,CMS:2017pet,Aaboud:2018qcu}.  For decays
dominantly to muon the limit is $0.87$ TeV \cite{Aaboud:2017qph}. The
search to $WW$ final states \cite{Aaboud:2018qcu} does not apply to
our model. Note again, that these searches apply for prompt decays.

Finally, the MoEDAL~\cite{Acharya:2014nyr} experiment currently
running at the LHC is also sensitive to highly-ionising
particles. Although designed to search for magnetic monopoles (see for
instance~\cite{Acharya:2017cio}), it also has sensitivity to
electrically-charged massive long-lived particles. Our states are
potentially detectable in MoEDAL if they are produced with low
velocities ($\beta<0.2$) and travel at least a meter. These would
complement the accessible region at ATLAS and CMS, usually targeting
$\beta>0.2$, as mentioned above.

In summary, limits on the scalars of our model can be derived from 
various existing searches. Background-free searches could be done 
for the parts of parameter space, where at least $H^{4+}$ is long-lived 
enough to leave a charged track in the detector. None of the current 
searches, however, covers exactly the final states from $H^{4+}$ decays. 
We can therefore give only rough estimates on the mass limits for 
our model at present. 

\section{Summary and discussion}

\begin{figure}[tbph]
\begin{centering}
\includegraphics[scale=0.6]{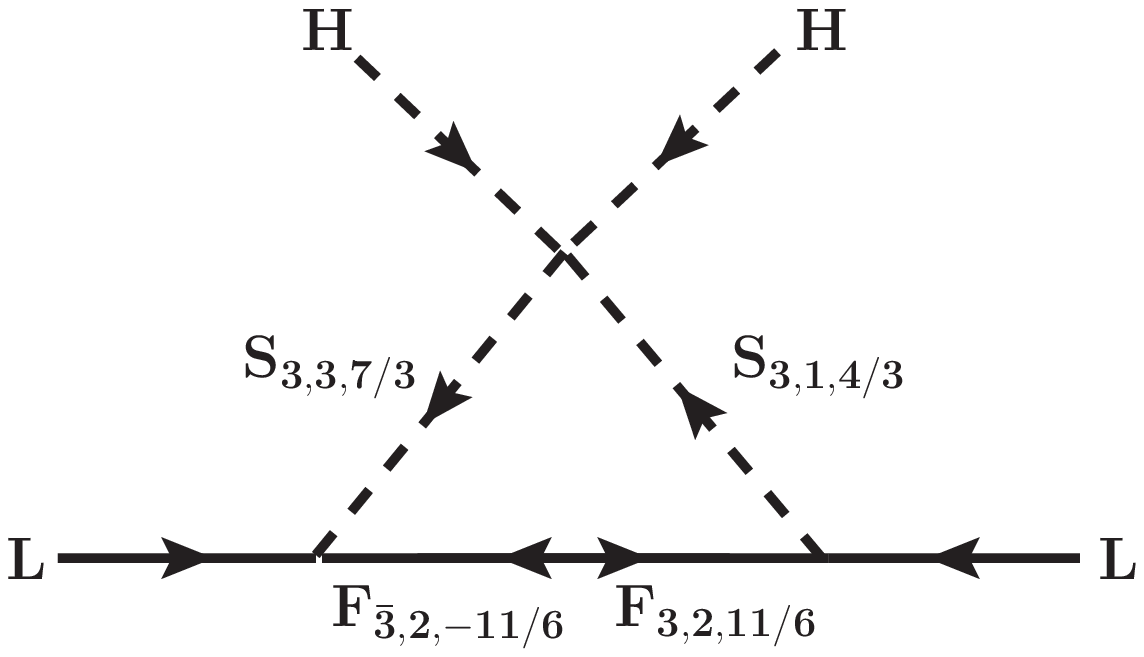}
\includegraphics[scale=0.6]{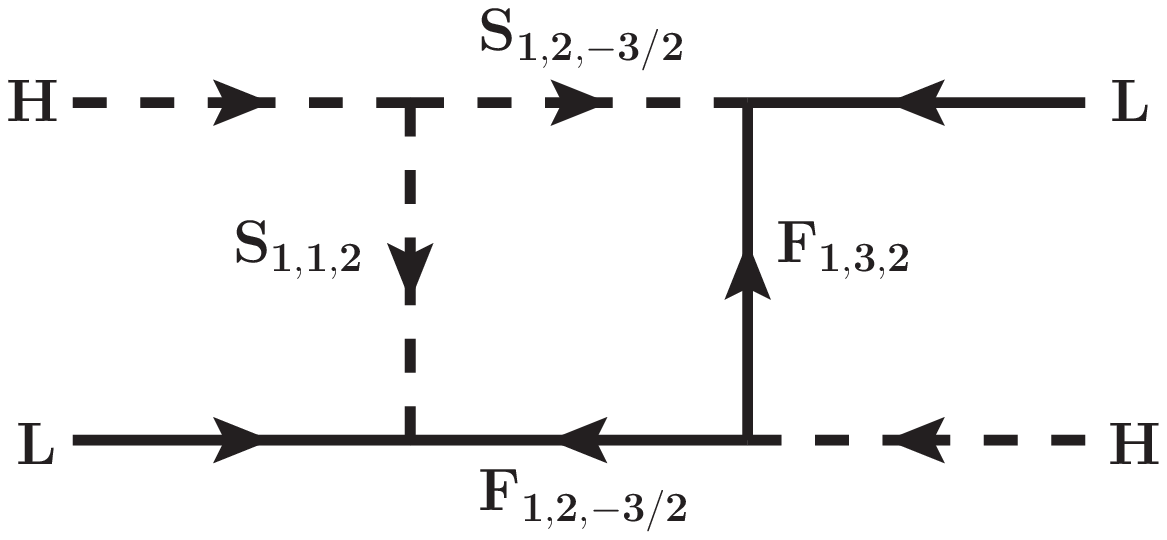}
\end{centering}
\protect\caption{\label{fig:vars} Two more variants of $d=5$ 1-loop models 
that lead to exotic LNV signals at the LHC. Left (model-II) a leptoquark 
variant, right (model-III) a variant based on a different one-loop 
neutrino mass topology. For a discussion see text.}
\end{figure}

We have discussed multi-lepton LNV final states at the LHC. We have
shown in one concrete 1-loop neutrino mass model that multi-lepton
signals can occur with experimentally interesting rates.  Moreover,
the smallness of the observed neutrino masses, together with the high
multiplicity of the final states, lead to tiny decay widths of the
multi-charged scalars, resulting in long charged tracks in large parts
of the available parameter space. This kind of exotic signals would be 
virtually background-free at the LHC (or a hypothetically future 
100 TeV $pp$-collider). Already with existing data, interesting bounds 
can be derived and we have discussed how to recast several searches in 
terms of the mass parameters of our model.

Before closing, we would like to discuss that the particular example
we studied in this paper is not the only (loop) model of neutrino
mass that leads to these exotic multi-lepton signals. Many other
variants following the same basic idea can be constructed.  Two
examples are shown in fig. (\ref{fig:vars}).

Model-II, see the diagram on the left of fig. (\ref{fig:vars}), is
based on the same topology as our proto-type model. However, here the
particles in the loop carry colour. The model has two different scalar
leptoquark states plus an exotic coloured fermion. Similar to our
proto-type model, the phenomenologically most interesting decays are
those of the scalar electro-weak triplet. This multiplet contains a
particle $S^{10/3+}_3$, which can decay either to $(3l^+)+j$ or
$l^++(2W^+)+j$. The LNV final state from pair-produced $S^{10/3+}_3$
will then consist of $(3l^\pm)l^{\mp}(2W^{\mp})(2j)$. Compared to the
$6l+2W$ signal of our proto-type model, one can thus have signals with
more jets and fewer charged leptons.\footnote{Leptonic decays of $W$
  add more leptons, but also missing energy to the observed final
  state.}  On the other hand, since the neutrino mass calculation for
this model and the multiplicity of the final state is very similar to
the ones discussed in this paper, one can expect the different states
of the $S_{3,3,7/3}$ to be long-lived, again, if they are the lighter
than $S_{3,1,4/3}$ and the exotic fermion. Cross sections for coloured
particles are larger than for coloured singlets at the LHC, so one can
expect even larger sensitivity in $pp$-colliders for such kind of model
variants.

Model-III, see the diagram on the right of fig. (\ref{fig:vars}), is a
model variant based on a different 1-loop topology. We have chosen to
show this example to demonstrate that models with multi-lepton final
states can arise from any loop topology. In this model, the particle 
with the largest charge is a triply charged fermion. The LNV multi-lepton 
final state for pair production of this particle will be 
$(3l^\pm)l^{\mp}(2W^{\mp})$, similar to model-II above, but with fewer 
jets. For multiply charged fermions, however, QED radiative corrections 
will always produce a small mass splitting between states with different 
electrical charges and thus one finds for a $F_{1,3,2}$ that 
roughly $m_{F^{3+}} - m_{F^{2+}} \sim 1.5$ GeV. The $F^{3+}$ will then 
decay to $F^{2+}+\pi^+$, unsuppressed by the small neutrino masses and 
this puts an upper limit on the decay length of $F^{3+}$. In this 
model variant one therefore does not expect charged tracks with 
LNV signals at the end of the tracks.

Multi-lepton signals are not limited to $d=5$ neutrino mass models.
Rather, using higher-dimensional operators to generate neutrino
masses, requires larger representations (with non-zero
hypercharge). At $d=7$ one finds at tree-level the ``BNT'' model
\cite{Babu:2009aq}.  This model uses a scalar $S_{1,4,3/2}$, which
contains a triply charged scalar. The decays of (pairs of) the
$S^{3+}$ of this model can lead to the final state
$(2l^{\pm})W^{\pm}(3W^{\mp})$, with only two leptons, but accompanied
by many jets from the hadronic decays of the $W$. The decays of
$S^{3+}$, however, are expected to be rather short
\cite{Arbelaez:2019cmj}.  For neutrino mass models at $d=7$ and
1-loop, multi-lepton signals will actually be the norm and not the
exception. This can be seen from the example models discussed in
\cite{Cepedello:2017lyo} and easily deduced from the list of all
possible $d=7$ 1-loop diagrams given in \cite{Cepedello:2017eqf} .

In summary, multi-lepton signals, as discussed in this paper, can
arise in a variety of neutrino mass models. Since these high
multiplicity final states should have very little background at the
LHC, dedicated searches have strong discovery potential. On top of
that, even less background is expected if the charged particles are
long-lived. None of the existing searches is optimized for the signals
we have discussed, but interesting limits can already be derived from
recasting current data.

\begin{acknowledgements}
M.~H. is supported by the Spanish grant FPA2017-85216-P (AEI/FEDER,
UE) and PROMETEO/2018/165 (Generalitat Valenciana).  G.C. acknowledges support from 
FONDECYT-Chile grant No. 3190051. J. C. H. is supported by Chile grant FONDECYT 
No. 1161463. G.C. and J. C. H. also acknowledge support from grant FONDECYT No. 1201673. 
C.A. is supported by FONDECYT-Chile grant No. 11180722 and ANID-Chile PIA/APOYO AFB 180002.

\end{acknowledgements}



\end{document}